\documentclass[fleqn,usenatbib]{mnras}
\usepackage{amsmath}	
\usepackage{mathptmx}
\usepackage{txfonts}
\usepackage[T1]{fontenc}
\usepackage{rotating}
\DeclareRobustCommand{\VAN}[3]{#2}
\let\VANthebibliography\thebibliography
\def\thebibliography{\DeclareRobustCommand{\VAN}[3]{##3}\VANthebibliography}

\usepackage{graphicx}	

\usepackage{float}
\usepackage{hyperref}

\title[First DST Views of Chromospheric Swirls]{First High-Resolution Observations of Chromospheric Swirls with the Dunn Solar Telescope}

\author[Vesa, Shetye, and Verwichte]{
Oana Vesa,$^{1}$\thanks{E-mail: ovesa@stanford.edu}
Juie Shetye,$^{2}$
and Erwin Verwichte$^{3}$
\\
$^{1}$W. W. Hansen Experimental Physics Laboratory, Stanford University, Stanford, CA 94305\\
$^{2}$Department of Astronomy, New Mexico State University, Las Cruces, New Mexico 88003\\
$^{3}$Centre for Fusion, Space and Astrophysics, University of Warwick, Coventry CV4 7AL, UK
}

\date{Accepted XXX. Received YYY; in original form ZZZ}

\pubyear{\the\year{}}
\graphicspath{{./Figures/} }

\begin{document}
\label{firstpage}
\pagerange{\pageref{firstpage}--\pageref{lastpage}}
\maketitle
\begin{abstract}
We present the first observations of chromospheric swirls using the Hydrogen-alpha Rapid Dynamics camera and Rapid Oscillations in the Solar Atmosphere imaging instruments at the Dunn Solar Telescope. These vortices contribute to heating and dynamics across the solar atmosphere. We analyze the morphology and evolution of 34 swirls and their cospatial bright points (BPs) from the photosphere to the mid-chromosphere. To examine swirl–BP interactions and temporal behavior, we use image segmentation, Fourier and spectral analysis, and local correlation tracking. The observed swirls have an average lifetime of 7.9\,$\pm$\,5\,min and diameter of 3.6\,$\pm$\,1\,Mm, with a positive correlation indicating smaller swirls tend to be short-lived. 76$\%$ are associated with a compact BP appearing 12\,s to 9\,min after swirl formation. Swirl motion is also closely linked to their BP(s) global motions. The swirls exhibit a mean angular speed of 0.04\,rad\,s$^{-1}$, radial speed of 17.7\,km\,s$^{-1}$, and period of 180\,s. We observe the formation of a spiral-shaped swirl driven by a BP interacting with a large photospheric vortex. The BP is dragged toward the vortex centre, after which the swirl forms. The BP undergoes changes in orientation and elongation that mirror the swirl's chromospheric development. A time lag of $-42.5$\,s between the sudden change in the BP's orientation and the peak of the swirl's intensity variation suggests torsional Alfvén waves may contribute to swirl evolution. Our results support a magnetic origin for swirls rooted in motions of photospheric BPs.
\end{abstract}

\begin{keywords}
Sun: atmosphere -- Sun: chromosphere -- Sun: photosphere
\end{keywords}



\section{Introduction}\label{sec:introduction}

Vortex motions of various spatiotemporal scales, driven by the dynamic interplay between the convective plasma and magnetic field, are observed across the solar atmosphere \citep[for a complete review, see][and references therein]{2023_Tziotziou}.
Because vortices are rooted in the top layer of the convective zone, they represent an integral part of the solar dynamo, coupling the subphotospheric flows and magnetic fields with the overlying atmospheric regions \citep{1985_Nordlund, 2012_Stein, 2012_Wedemeyer-Bohm_etal_Nature}.
We focus on small-scale vortex flows called chromospheric swirls, commonly observed in the quiet Sun chromosphere \citep{2009_WedemeyerBohm_RouppevanderVoort} and resolved in numerical simulations \citep{2010_Carlsson, 2021_Yadav}.
These structures are thought to transport significant amounts of energy into the upper atmosphere, serving as conduits for heating mechanisms through Alfvén pulses and wave dissipation \citep{2012_Wedemeyer-Bohm_etal_Nature,2019_Liu, 2021_Battaglia_CaniveteCuissa_Calvo_etal}.

Chromospheric swirls are magnetic, rotating structures rooted in the convective zone that arise from intergranular flows \citep{2012_Wedemeyer-Bohm_etal_Nature}.
Illustrative examples can be found in \citet[see Fig.\,1]{2014_Wedemeyer_Steiner} and \citet[see Fig.\,33]{2023_Tziotziou}.
They have visible footpoints on the photosphere in the form of bright points (BPs) located in intergranular lanes, which possess kG magnetic field strength \citep{2008_Bonet_etal, deWijn_2009, Keys_2019}.
Above these photospheric flows that are a direct result of the conservation of angular momentum, the twisting magnetic field and plasma can spiral in an upward trajectory, producing a swirl-like imprint in the chromosphere or ``magnetic tornado'' \citep{2012_Kitiashvili, 2014_Wedemeyer_Steiner}.
Observations and numerical simulations show that BPs have matching rotary motions to the overlying chromospheric swirl, further cementing their connection \citep{2011_Shelyag, 2019_Shetye_Verwichte}.

Chromospheric swirls were identified by \citeauthor{2009_WedemeyerBohm_RouppevanderVoort} in high-resolution spectroscopic narrowband imaging of \ion{Ca}{ii}\,8542\,{\AA} with the CRisp Imaging SpectroPolarimeter \citep[CRISP;][]{2008_Scharmer_Narayan_Hillberg_CRISP} at the Swedish 1-m Solar Telescope \citep[SST;][]{2003_Scharmer_SST}.
Subsequent ground and space-based observations identified them in other chromospheric diagnostics (i.e, H$\alpha$\,6563\,{\AA} and \ion{Mg}{ii k}\,2796\,{\AA}) as swirling intensity imprints that appear as dark or bright rings, spiral arms, ring fragments, and arcs \cite[see][]{2013_Wedemeyer_etal, 2016_Park_Tsiropoula_Kontogiannis_Tziotziou_etal, 2018_Tziotziou_Tsiropoula_etal, 2019_Shetye_Verwichte}.
Detection of chromospheric swirls has been primarily by eye; however, in recent years, automated methods have emerged to identify these vortices \citep{2017_Kato_Wedemeyer, 2021_Dakanalis_etal, 2022_Dakanalis}.
Still, there exists a discrepancy between numerically simulated \citep[N = 1 swirl Mm$^{-2}$ min$^{-1}$;][]{2017_Kato_Wedemeyer} and observed \citep[N $\simeq$ 10$^{-2}$ swirl Mm$^{-2}$ min$^{-1}$;][]{2022_Dakanalis} swirls.

Swirls have observed diameters between 0.5--3\,Mm and average lifetimes of 10\,min, with the range of statistical properties restricted by current instrumental resolution \citep{2009_WedemeyerBohm_RouppevanderVoort,2022_Dakanalis}.
While chromospheric swirls are typically observed as isolated features, \citet{2018_Tziotziou_Tsiropoula_etal, 2019_Tziotziou_Tsiropoula_Kontogiannis_PII, 2020_Tziotziou_Tsiropoula_Kontogiannis_PIII} identified a ``small-scale tornado'' in H$\alpha$ with a diameter of 4.4\,Mm and lifetime of 1.7\,hr that consisted of multiple smaller-scale swirls that intermittently appeared and disappeared throughout the observational period and exhibited clear interactions with neighboring fibrils.
The first spectropolarimetric analysis of a pair of co-rotating, interacting chromospheric swirls by \citet{2020_Murabito_Shetye_etal} confirmed that these structures can be connected locally in the chromosphere and provided direct evidence of magnetic coupling across atmospheric layers.

Swirls exhibit rotating, swaying, and radial expansion motions \citep{2019_Tziotziou_Tsiropoula_Kontogiannis_PII}.
Analysis of Dopplergrams indicates fast central up flows in the range of 2--13\,km\,s$^{-1}$ and radial expansions of the spiral arms reaching projected velocities between 4--20\,km\,s$^{-1}$ \citep{2012_Wedemeyer-Bohm_etal_Nature, 2018_Tziotziou_Tsiropoula_etal, 2019_Shetye_Verwichte}.
These central upflows revealed temperature increases, highlighting the role of swirls as energy agents \citep{2016_Park_Tsiropoula_Kontogiannis_Tziotziou_etal}.
Additionally, analysis by \citet{2020_Murabito_Shetye_etal} suggests that a swirl's rotation may be apparent rather than real due to the superposition of slow magnetoacoustic modes.

We present a morphological study of 34 chromospheric swirls observed for the first time using high-resolution, high-cadence imaging observations from the Dunn Solar Telescope (DST)\footnote{\href{https://sunspot.solar/about/}{https://sunspot.solar/about/}}, tracking their formation, evolution, and dynamics from the photosphere to mid-chromosphere.
This work aims to provide insight into how BP dynamics correlate with chromospheric swirls, expanding upon previous studies \citep[e.g.][]{2019_Shetye_Verwichte}.
Additionally, the broader H$\alpha$  bandpass filter employed in this work provides a different visual perspective of swirl evolution, as the intensity contributions from the H$\alpha$ wings are not isolated as in previous works \citep[i.e.,][]{2018_Tziotziou_Tsiropoula_etal}.
This study will provide a baseline for similar investigations with the Visible Broadband Imager \citep[VBI;][]{2021_Woger_VBI} on the 4-m Daniel K. Inouye Solar Telescope \citep[DKIST;][]{2020_Rimmele_DKIST}.

\begin{figure*}
    \centering
\includegraphics[width=\textwidth]{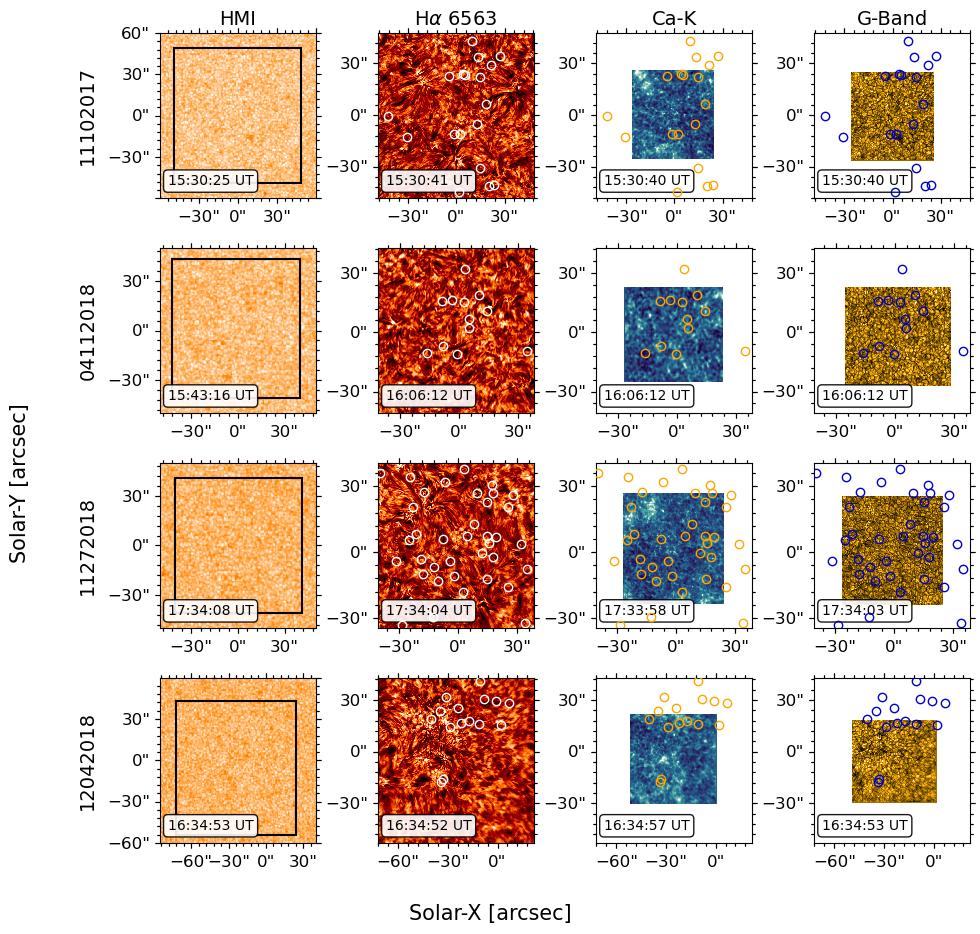}
    \caption{Quiet-Sun observations on 10 November 2017 (DS1), 11 April 2018 (DS2), 27 November 2018 (DS3), and 04 December 2018 (DS4). (First Panel): The nearest in-time HMI continuum image; no cotemporal HMI data were available for DS2. (Second Panel): HARDcam's H$\alpha$\,6563\,{\AA}, with the field of view marked by the black box in the HMI panels. (Third -- Fourth Panel): ROSA Ca-K and G-Band are aligned to the same field of view as H$\alpha$. Circles mark the approximate locations of the 85 visually identified chromospheric swirl candidates. Not all candidates were observed within the ROSA field of view.}
\label{fig:dataset_reference_images}
\end{figure*}

The paper is structured as follows.
The DST observations and the chromospheric swirl detection method are described in Section\,\ref{sec:observations}. 
Sections\,\ref{sec:photospheric_dynamics}\,--\,\ref{sec:rotational_dynamics} address the properties of the chromospheric swirls and their underlying photospheric dynamics.
Section\,\ref{sec:swirl_evolution} presents a case study of the formation of a large spiral-shaped swirl by the interaction of its footpoint with a large, persistent photospheric vortex.
The discussion and conclusion follow in Sections\,\ref{sec:discussion_and_conclusions}\,--\,\ref{sec:conclusions}.
Appendix \ref{sec:observation_table} and \ref{sec:swirl_properties} describe the properties of the observations and the surveyed chromospheric swirls.

\section{Observations} \label{sec:observations}

\subsection{Data} \label{subsec:data_andswirlproperties}

\begin{figure}
\includegraphics[width=\columnwidth]{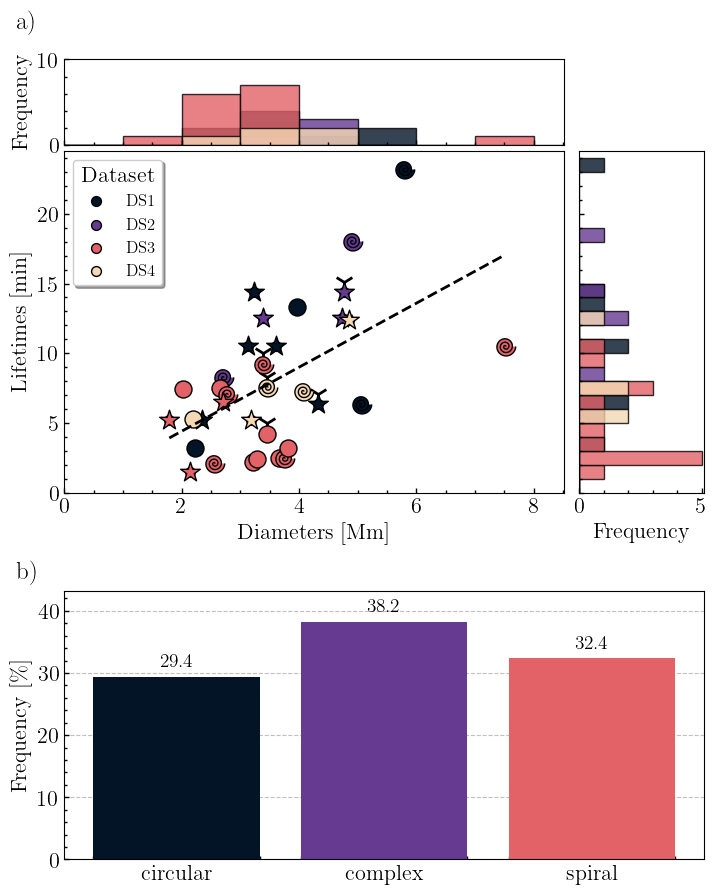}
    \caption{Properties of the observed 34 chromospheric swirls in H$\alpha$. a) Moderate positive correlation between the estimated diameters and lifetimes (Pearson coefficient of 0.55). Markers indicate swirl morphology: stars for complex, circles for circular, and spirals for spiral-shaped swirls. Downward triangle markers represent the minimum lifetimes for swirls that began forming before observations. b) Percentage distribution of H$\alpha$ intensity imprints by morphological category (circular, complex, or spiral).}\label{fig:diameter_lifetime_histogram_scatter}
\end{figure}

We present high-resolution, high-cadence observations from the Rapid Oscillations in the Solar Atmosphere \citep[ROSA;][]{2010_ROSA_instrumentation_paper} and Hydrogen-Alpha Rapid Dynamics Camera \citep[HARDcam;][]{2012_HARDcam} imaging instruments installed at DST in Sunspot, New Mexico.
These instruments use the DST's high-order adaptive optics system \citep{2004_Rimmele_Richards_AO}.
We focus on four quiet-Sun near disk centre datasets taken on 10 November 2017, 11 April 2018, 27 November 2018, and 04 December 2018\footnote{Datasets are publicly available on the Sunspot Solar Observatory Consortium (SSOC) Data Archive: \href{http://ssoc.nmsu.edu/}{http://ssoc.nmsu.edu/}}.
These datasets will be referred to herein as DS1, DS2, DS3, and DS4.

HARDcam observations used a 0.25\,{\AA} bandpass filter centred on H$\alpha$\,6562.8\,{\AA} with a spatial sampling of 0{\farcs}083 per pixel.
The observations range from 36 to 102 min with reconstructed cadences between 2.5 and 3.2\,s.
ROSA observed a smaller, co-spatial field of view in three broadband filters with a spatial sampling of 0{\farcs}06 per pixel: (1) G-Band centred on 4305.5\,{\AA} with a 0.9\,{\AA} bandpass filter; (2) blue continuum centred on 4170\,{\AA} with a 52\,{\AA} bandpass filter; and (3) Ca-K centred on 3933.7\,{\AA} with a 1\,{\AA} bandpass filter.
These datasets range from 40.9 to 82.1\,min, except for DS4 Ca-K, which lasts 7.6\,min.
G-Band and 4170\,{\AA} have the same reconstructed cadence: 2.1\,s for DS1 and DS3 and 4.2\,s for DS2 and DS4.
Ca-K reconstructed cadences range from 8.4 to 12.6\,s.
Dataset properties are available in Table\,\ref{tab:dataset_overview}.
Although the 4170\,{\AA} filter is described for completeness, we omit it from analysis due to its similarity to the G-Band filter.

Calibration data acquired during the science observations were used to provide initial co-alignment of the channels.
The datasets were also speckle reconstructed with the Kiepenheuer-Institut Speckle Interferometry Package \citep[KISIP;][]{2008_Woger_KISIP}, producing diffraction-limited images to match the DST's angular resolution ($\approx$\,0.14\arcsec) and destretched to remove atmospheric distortions.
Co-alignment among the channels, which can be seen in Fig.\,\ref{fig:alignment_check}, was refined using a 2D cross-correlation technique to match bright, large-scale features present throughout the atmosphere, with ROSA channels first degraded to match the resolution of HARDcam.

The co-aligned datasets are shown in Fig.\,\ref{fig:dataset_reference_images}, where the crosses indicate visually identified chromospheric swirl candidates.
These observations cover the low photosphere to the middle chromosphere: H$\alpha$ samples the middle chromosphere \citep[around 1500\,km;][]{1981_Vernazza_Avrett_Loeser}; Ca-K is representative of the upper photosphere to the low chromosphere \citep[around 1300\,km;][]{1969_Beebe_Johnson}; and G-Band and 4170\,{\AA} sample the low photosphere \citep[100\,km and 25\,km, respectively;][]{2012_Jess_Shelyag}.

\subsection{Swirl Properties and Statistics} \label{subsec:overview_swirls}

\begin{figure*}
    \centering
\includegraphics[width=0.65\textwidth]{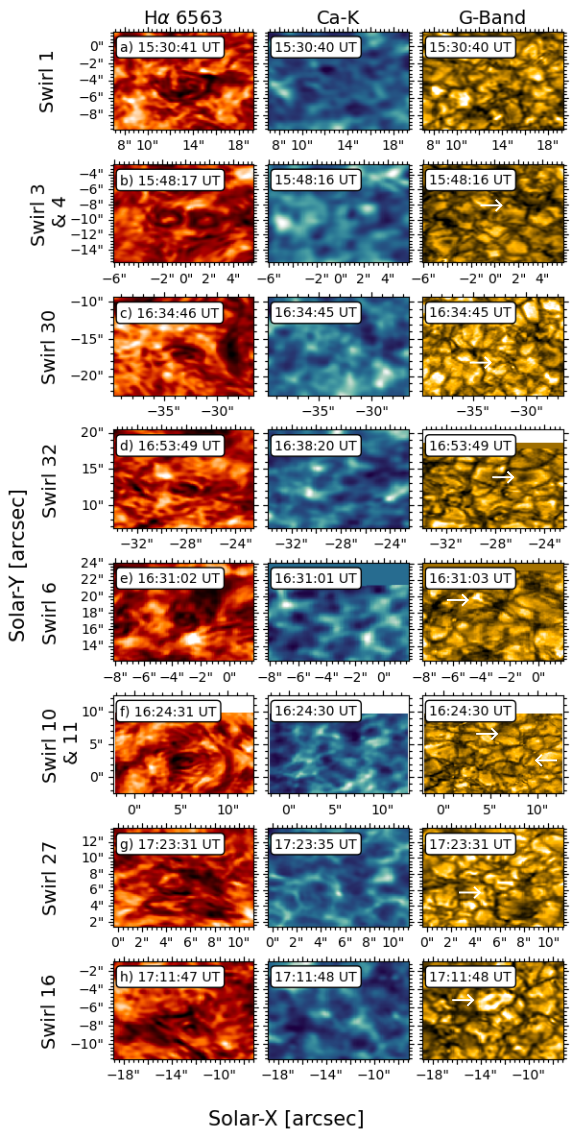}
    \caption{H$\alpha$, Ca-K, and G-Band intensity images of eight chromospheric swirls, corresponding to those in Appendix\,\ref{sec:swirl_properties}. These examples illustrate the diversity in morphology, temporal evolution, and underlying photospheric dynamics of chromospheric swirls. White arrows in the photospheric sequences highlight features within granulation cells (e.g., dark dots and dark lanes) that emerge contemporaneously with swirl formation and evolution.}
\label{fig:select_swirling_examples}
\end{figure*}

We visually detected 85 chromospheric swirls candidates in H$\alpha$: 18 in DS1, 13 in DS2, 38 in DS3, and 15 in DS4.
Similarly to \citet{2019_Shetye_Verwichte}, we require candidates to exhibit a clear spiral, ring, or circular intensity pattern. 
Because we aim to investigate the evolution and formation of chromospheric swirls, we excluded swirls outside the G-Band field of view due to a lack of photospheric information (see Fig.\,\ref{fig:dataset_reference_images}) and swirls that could not be properly analyzed due to unstable seeing conditions. 
However, we did not disqualify swirls lacking a cospatial BP in the photospheric G-Band.
Using these criteria, we identified 34 chromospheric swirl candidates that could be tracked from the low photosphere (G-Band) to the mid-chromosphere (H$\alpha$), which are seen in Appendix\,\ref{sec:swirl_properties}.

Fig.\,\ref{fig:select_swirling_examples} presents intensity snapshots of eight chromospheric swirls, highlighting their diverse characteristics.
Although their diverse appearances are seen in H$\alpha$, we did not detect any swirl-like structures in the corresponding Ca-K sequences.
This is likely because the Ca-K broadband filter is contaminated with signals from both the upper photosphere and low chromosphere \citep{2013_Morton}.
The Ca-K intensity images show various bright spots that are persistent or transient during the swirls' evolution.
We find no one-to-one correspondence between the bright spots in Ca-K and those in H$\alpha$ and no one-to-one correspondence to the G-Band BPs, in agreement with \citet{2013_Keys_Mathioudakis}.

Although most swirls appear to form independently and are shaped by their surrounding environment, we identify three pairs of ``twin swirls''.
These swirls evolve in proximity to and interact with one another.
So far, there has only been a single report of a pair of interacting swirls by \citet{2020_Murabito_Shetye_etal}.
Examples of twin swirls are shown in Fig.\,\ref{fig:select_swirling_examples}.
We note that Swirl 22 has a visible ``partner''; however, since the partner does not appear to display any swirling motion, it was excluded.

Lifetimes and diameters were estimated by assessing H$\alpha$ intensity signatures.
We estimated lifetimes by visually tracking the swirl, similarly to \cite{2019_Shetye_Verwichte}.
For four swirls, lifetimes likely began before the H$\alpha$ time series; thus, their start times were set to match the time series and noted in Appendix\,\ref{sec:swirl_properties}.
We calculated event diameters using the largest circular extent that enclosed the full H$\alpha$ imprint at the swirl's most evolved state.

The observed swirls have an average lifetime of 7.9\,$\pm$\,5\,min and average diameter of 3.6\,$\pm$\,1\,Mm.
Fig.\,\ref{fig:diameter_lifetime_histogram_scatter}a illustrates a moderate positive correlation between the variables (Pearson coefficient of 0.55).
This correlation suggests that smaller swirls tend to have shorter lifetimes than larger swirls.

We classified the swirls into three morphological categories based on their H$\alpha$ intensity imprint: circular, spiral, or complex.
A swirl was classified as complex if it did not cleanly fall into the circular and spiral categories or demonstrated changing visual appearances (e.g., spiral to circular). 
As seen in Fig.\,\ref{fig:diameter_lifetime_histogram_scatter}b, we find that 29$\%$ displayed a circular pattern, 32$\%$ showed a spiral pattern, and 38$\%$ were classified as complex.
From the complex category, we note that 14$\%$ demonstrated changing visual imprints.
The swirls that showed this were two pairs of twin swirls (spiral to circular: Swirls 3 and 4 and Swirls 12 and 13) and Swirl 33 (complex to spiral).

\section{Photospheric Dynamics} \label{sec:photospheric_dynamics}
\begin{table}
\caption{Summary of average BP and swirl properties and their correlations.}
    \label{tab:BP_stats_correlations}
    \centering
    \small
    \begin{tabular}{c|c}
    \hline
    \hline
    \textbf{BP Properties} & \textbf{Percentage (Count)} \\
    \hline
     Average Lifetime [min] &  (17.1\,$\pm$\,10)   \\
     Average Diameter [km]   & (175\,$\pm$\,70)\\
     Average Translational Speed [km\,s$^{-1}$] & (4.5\,$\pm$\,3)\\
     Average Area  [km$^{2}$] & (27,000\,$\pm$\,19,000)\\
     RMS Angular Speed [rad\,s$^{-1}$] & (0.06\,$\pm$\,0.0009)\\
    \hline
    Linear Trajectory & 29\,$\%$ (12)\\
    Semicircular Trajectory & 61\,$\%$ (25) \\
    Serpentine Trajectory & 10\,$\%$ (4)\\
    \hline
    CW Turning Direction & 46\,$\%$ (19)\\
    CCW Turning Direction & 23\,$\%$ (11) \\
    Collinear Turning Direction & 23\,$\%$ (11) \\
    \hline
    Dominant CW Orientation & 42\,$\%$ (17) \\
    Dominant CCW Orientation & 59\,$\%$ (24) \\
    \hline
    \hline 
    \textbf{Swirl Properties} & \textbf{Percentage (Count)} \\
    \hline
    Average Lifetime [min] & (7.9\,$\pm$\,5) \\
    Average Diameter [Mm] & (3.6\,$\pm$\,1) \\
    Average Area [Mm$^2$] & (10.2\,$\pm$\,6) \\
    Median Angular Speed [rad\,s$^{-1}$] & (0.04; IQR: 0.024-0.092) \\
    Median Radial Speed [km\,s$^{-1}$] & (17.7; IQR: 7.8-8.6) \\
    Average Period [s] & (180.0\,$\pm$\,114) \\
    \hline
    CW Rotation & 35\,$\%$ (12) \\
    CCW Rotation & 47\,$\%$ (16) \\
    Changing Rotation & 18\,$\%$ (6) \\
    \hline
    \hline
    \textbf{BP-Swirl Correlations}  & \textbf{Percentage (Count)}  \\
    \hline
    Orientation Matches Trajectory & 39\,$\%$ (16) \\
    Orientation Matches Swirl Rotation & 49\,$\%$ (20) \\
    Trajectory Matches Swirl Rotation & 37\,$\%$ (15) \\
    Orientation Does Not Match Swirl Rotation & 12\,$\%$ (5) \\
    Trajectory Does Not Match Swirl Rotation & 15\,$\%$ (6) \\
    Neither Match Swirl Rotation & 15\,$\%$ (6) \\
    \hline
    Appear Before Swirl Forms & 34\,$\%$ (14) \\
    Appear After Swirl Forms & 69\,$\%$ (27) \\
    Disappear After Swirl Disappears & 27\,$\%$ (11) \\
    Disappear During/Before Swirl Disappears & 72\,$\%$ (30) \\
    \hline
    \end{tabular}   
\end{table}

\begin{figure*}
     \centering
\includegraphics[width=\textwidth]{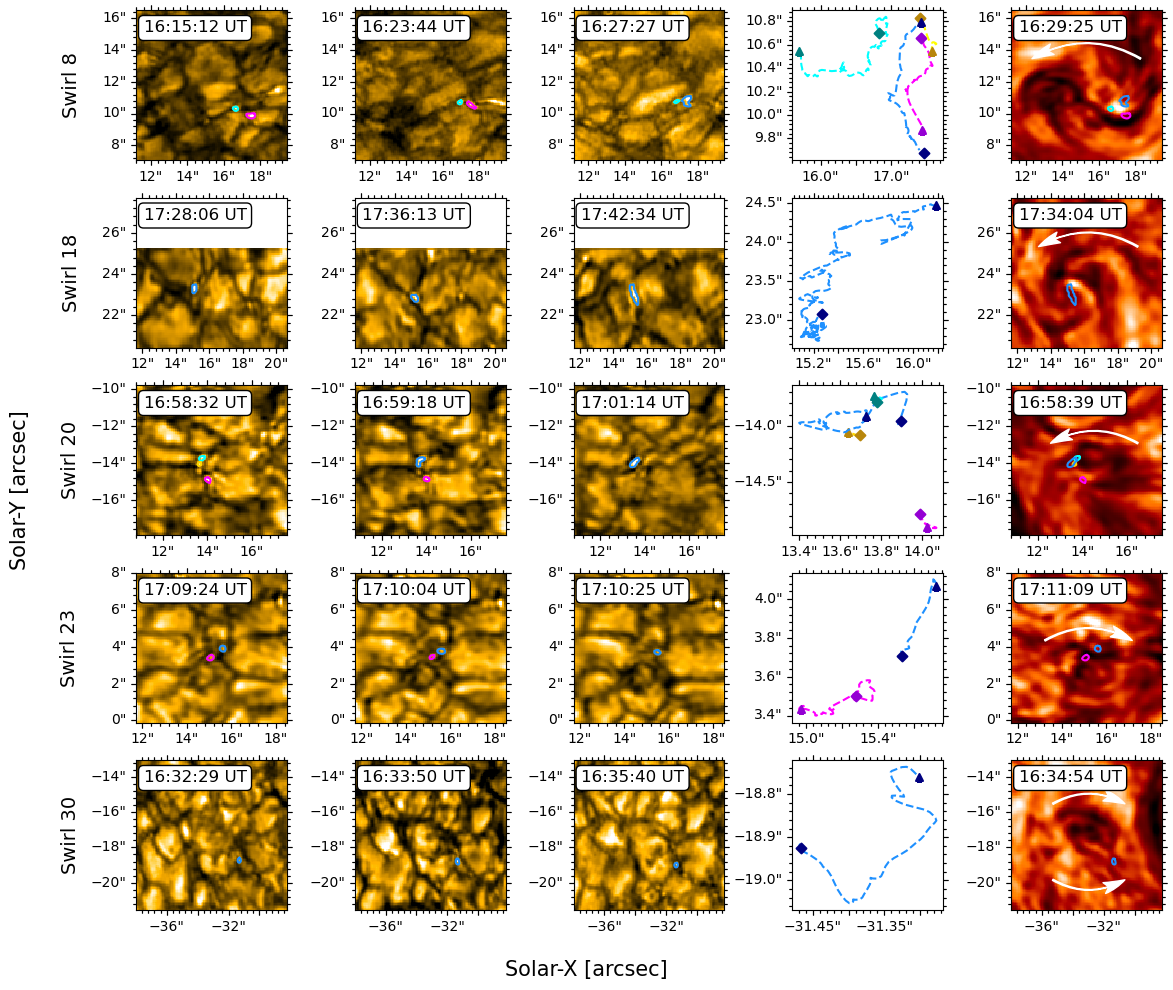}
    \caption{Examples of BP dynamics and their associated chromospheric swirls. (First -- Third Panel): Co-spatial G-Band snapshots showing the temporal evolution of the BP(s). (Fourth Panel): Zoomed-in view of BP trajectories derived from the tracking procedure. Trajectories are color-coded to match the BPs in the G-Band sequences. Triangle markers denote the start of each BP's evolution, and diamond markers denote the end. (Fifth Panel): Corresponding chromospheric swirls in H$\alpha$ with BP contours marking their footpoint(s). White arrows indicate the overall rotation of the chromospheric swirl (see Subsection\,\ref{subsec:chromosphere_rotational_dynamics}).}
    \label{fig:MBP_tracking_example}
\end{figure*}

We investigate the underlying photospheric dynamics and footpoints of swirls by tracking the temporal evolution of all their surrounding cospatial BPs in G-Band \citep[e.g.,][]{Nisenson_2003, Utz_2010, Chitta_2012, 2013_Keys_Mathioudakis}.
We exclude BPs located too close to the image edge and those with weak intensities and short lifetimes from the tracking analysis.
BPs were isolated by applying a circular mask around their centre, the diameter of which was slightly larger than the swirl.
Local thresholding and contour-finding techniques were applied to perform image segmentation using Python's \textit{scikit-image} and manually track the BPs until they disappeared, merged, or fragmented.
The results of this method are presented in Fig.\,\ref{fig:MBP_tracking_example}.

We derived the intensity-weighted centroid coordinates, mean intensity, equivalent diameter, orientation, and area throughout the BP's lifetime\,\footnote{Information about the derived properties can be found at \href{https://scikit-image.org/docs/stable/api/skimage.measure.html\#skimage.measure.regionprops}{scikit-image}.}. 
Translational and angular speed were also computed.
We analyzed the BP's trajectory curvature by computing the signed 2D cross-product of successive intensity-weighted centroid coordinates to classify the overall turning direction (clockwise, counterclockwise, or collinear) and categorize the trajectory path (linear, semicircular, or serpentine).
A serpentine trajectory is defined as when the curvature of the BP's trajectory changes frequently and significantly.
The circular mean of the derived orientation angles was also calculated to determine the BP's dominant angular motion.
Key results are summarized in Table\,\ref{tab:BP_stats_correlations}.

We tracked 41 BPs, including newly formed BPs produced through fragmentation and mergers.
Our sample set has a mean diameter of 175.3\,$\pm$\,65\,km, mean area of 26,971\,$\pm$\,19,338\,km$^{2}$, mean lifetime of 17.1\,$\pm$\,13\,min, mean translational speed of 4.5\,$\pm$\,3\,km\,s$^{-1}$, and RMS angular speed of 0.06\,$\pm$\,0.0009\,rad\,s$^{-1}$, which are similar to previously observed properties \citep[e.g.,][]{Abramenko_2010, Crockett_2010, Utz_2010, Keys_2011, Berrios_Saavedra_2022}.
The breakdown of their trajectories, as seen in Fig.\,\ref{fig:MBP_tracking_example}, are as follows: 25 exhibit a semicircular trajectory (e.g., Swirl 30's BP); 12 exhibit a linear trajectory (e.g., Swirl 23's BP); and 4 exhibit a serpentine trajectory (e.g., Swirl 18's BP).
These trajectories align with those reported in works by \citet{2011_MansoSainz_etal}, \citet{Abramenko_2011}, and \citet{Chitta_2012}.
\citet{2011_MansoSainz_etal} identified the smaller closed circular loops within the BP's overall trajectory (e.g., Swirl 18 and Swirl 23) as whirlpool motions associated with convective downdrafts similar to those observed by \citet{2008_Bonet_etal}.

The BPs preferentially show an overall clockwise turning direction in their trajectory (42$\%$), with an equal frequency exhibiting an overall counterclockwise and collinear turning direction (27$\%$).
Computation of the BPs' dominant derived orientation angle shows a modest difference between clockwise (42$\%$) and counterclockwise (59$\%$) motion.

We observe 76$\%$ of swirls have at least one cospatial BP, whereas 23$\%$ do not have a BP that can be detected as a G-Band intensity enhancement.
From the former, approximately 58$\%$ are associated with a single BP, and 17$\%$ are associated with BP groups.
Swirls appear more frequently with compact, quasi-circular than elongated BPs.
We also observe that 26$\%$ of swirls are associated with BPs that exhibit visible changes in morphology (e.g., circular to elongated), and 26$\%$ of swirls are associated with BPs that exhibit merging or fragmentation.
However, three swirls associated with multiple BPs showed no direct visual interaction among the BPs themselves.
Swirl 23 shows two BPs appearing to collide, but no obvious merging was detected in the photosphere, nor any observable effect on the swirl's H$\alpha$ intensity imprint.
Examples of various BP dynamics are shown in Fig.\,\ref{fig:MBP_tracking_example}.

In addition to the observed BP dynamics, the swirls are associated with changing appearances of the granules surrounding their footpoints, which coincide with their formation and evolution.
This includes an exploding granule \citep{1968_Carlier_etal} as in Fig.\,\ref{fig:select_swirling_examples}h whose outline is seen effecting the swirl's formation in the mid-chromosphere; lanes of leading bright rims and trailing dark edges indicative of horizontally oriented vortex tubes \citep{2010_Steiner_etal} as Swirl 6 evolves in Fig.\,\ref{fig:select_swirling_examples}e; dark dots \citep{1979_Kitai_Kawaguchi} as in Fig.\,\ref{fig:select_swirling_examples}b and d; a bright dot as in Fig.\,\ref{fig:select_swirling_examples}b which also corresponds to a noticeable bright spot in the mid-chromosphere; and developing and moving dark lanes that sometimes fragment the granule as in Fig.\,\ref{fig:select_swirling_examples}c, f, and g.
All of these are various types of strong downflows \citep{1999_Hirzberger_Bonet_Vazquez_Hanslmeier} and related to emerging magnetic fields \citep[for more details, see][]{2012_Palacios_etal,2020_Fischer_etal,2020Guglielmino_Salvo_etal}.

\section{Rotational Dynamics} \label{sec:rotational_dynamics}

\subsection{Chromospheric Swirls}\label{subsec:chromosphere_rotational_dynamics}
Chromospheric swirls are expected to show swaying motions, rotation, and radial expansion \citep[see][]{2014_Wedemeyer_Steiner}.
We explore rotational dynamics by constructing polar grids centreed on each swirl. 
The rotation is extracted by examining the intensity time-angle plots at different radial distances from the swirl centre. 
Radial expansion is investigated by examining the angularly integrated intensity as a function of radius and time, similarly to \citet{2018_Tziotziou_Tsiropoula_etal,2019_Shetye_Verwichte,2020_Murabito_Shetye_etal}.

\begin{figure*}
    \centering
\includegraphics[width=0.49\textwidth]{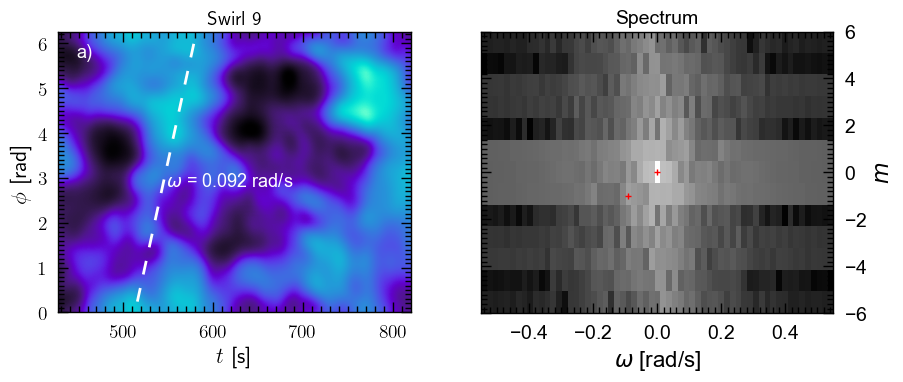}
\includegraphics[width=0.49\textwidth]{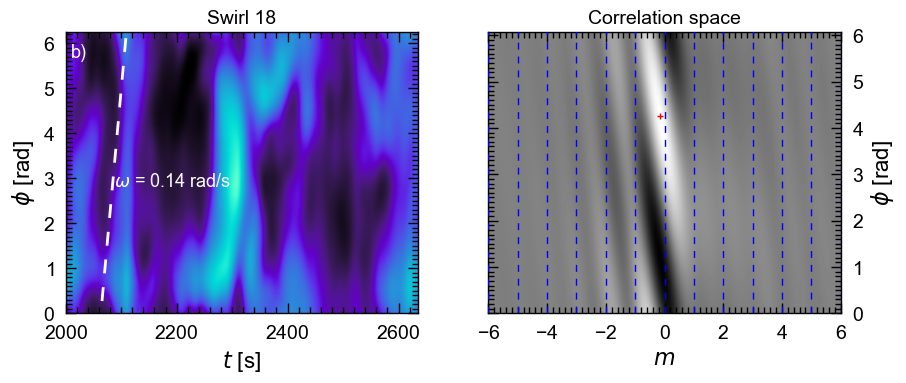}
\includegraphics[width=0.49\textwidth]{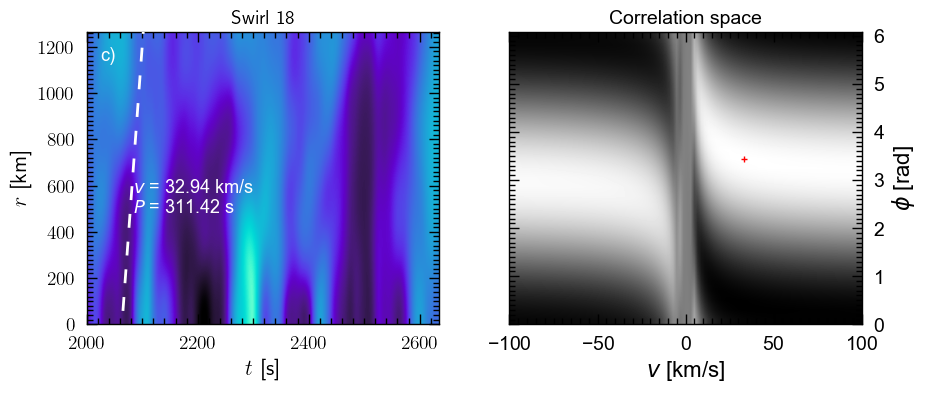}
\includegraphics[width=0.49\textwidth]{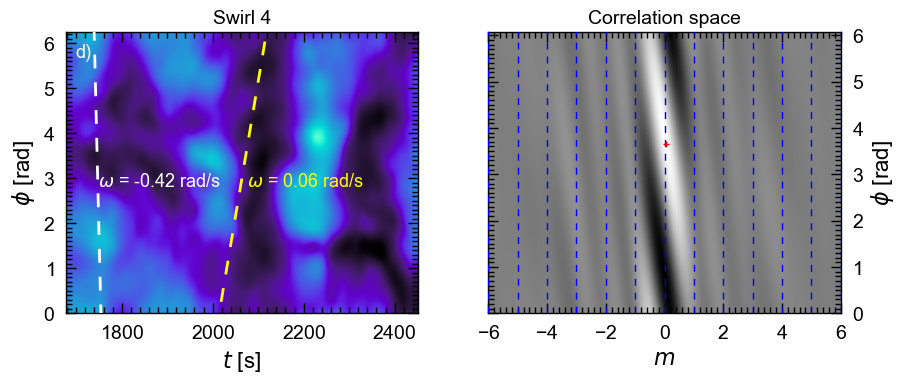}
\includegraphics[width=0.49\textwidth]{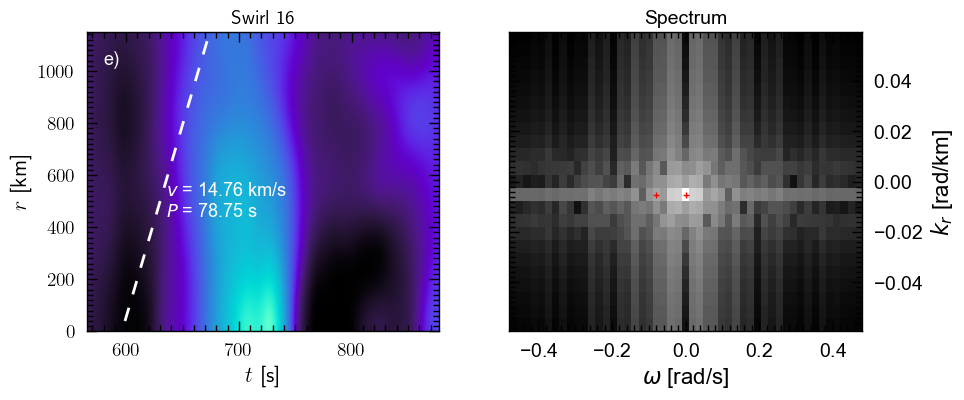}
    \caption{Azimuthal time-distance plots and radial distance-time plots centreed on a polar coordinate system using the method described in Subsection\,\ref{subsec:chromosphere_rotational_dynamics}. Correlation space and spectrum maps are also shown. a)--c) Examples where the automated methods fit the rotational behavior or radial expansion well. d)--e) Examples where the automated methods did not fit the rotational behavior or radial expansion well. A positive (negative) angular speed indicates counterclockwise (clockwise) motion. Likewise, a positive (negative) radial speed indicates outward (inward) propagation. The white (yellow) dashed lines show the results from the automated techniques (manual fit).}
\label{fig:rotational_dynamics_figure}
\end{figure*}

To minimize manual fitting, we employ automated tools to quantify rotational and wave propagation behavior \citep{2020_Murabito_Shetye_etal}. 
Two complementary techniques are used: spectral and correlation analysis. 
In spectral analysis, 2D Fourier spectra are derived from maps of azimuthal angle ($\phi$) versus time and radial distance ($r$) versus time. 
We are interested in integer azimuthal wavenumbers ($m$), and the rotation frequency is $\omega_{0}/m$.

In the correlation analysis, we apply the following procedure to the two types of maps. 
We first determine the most significant frequency ($\omega_{0}$). 
For the time-angle map, we maximize the correlation with a pattern of the form $\sin(m\phi - \omega_{0}t + \phi_{0})$ in the parameter space of $m$ and phase ($\phi_{0}$).
For the radius-time map, we maximize correlation with the pattern $\sin(\omega_0(t\!-\!r/v_{r}) + \phi_{0})$ in the parameter space of radial speed ($v_r$) and phase.
In both analysis methods, rotation is not established if no significant peaks are found at a non-integer value of $m$.
To analyze the full rotational behavior, we fit near the swirl centre (within 0.24--0.32\,Mm from the swirl centre) and near the swirl arms (dependent on swirl diameter). 
If the automated methods fail, we manually fit the rotational bands, if present. 

Examples where the methods fit the rotational behavior well are shown in Fig.\,\ref{fig:rotational_dynamics_figure}a--c.
In Fig.\,\ref{fig:rotational_dynamics_figure}a, the azimuthal time-distance plot of Swirl 9, measured within 0.81\,Mm from the centre, reveals an angular speed of $\omega = 0.092$\,rad\,s$^{-1}$ and azimuthal wavenumber $m = -1$ from spectral analysis.
Fig.\,\ref{fig:rotational_dynamics_figure}b shows a similar figure for Swirl 18, measured within 0.32\,Mm of the centre, with a fit of $\omega = 0.136$\,rad\,s$^{-1}$ and $m\,\simeq\,0$ using correlation analysis.
An example of the radial distance versus time is shown in Fig.\,\ref{fig:rotational_dynamics_figure}c for Swirl 18 using spectral analysis, where radial speed $v_r = 32.94$\,km\,s$^{-1}$ and period $P = 311$\,s.

However, the techniques fail in certain situations.
For instance, Swirl 4's behavior in Fig.\,\ref{fig:rotational_dynamics_figure}d was disrupted by its twin swirl at the beginning, and automated methods produce a poor fit with $\omega = -0.42$\,rad\,s$^{-1}$ and $m\,\simeq\,0$. 
A manual fit of $\omega = 0.06$\,rad\,s$^{-1}$ produces a better match.
Swirl 16 is another example (Fig.\,\ref{fig:rotational_dynamics_figure}e), where its dynamics are disrupted by the neighboring exploding granule, resulting in no clear radial speed signal.

Overall, we fit the rotational dynamics of 56$\%$ of swirls with the automated methods, partially fit 18$\%$ (e.g., the region near the swirl centre was fit well but not near the swirl arms), and manually fit 26$\%$.
For the radial speeds, 94$\%$ of swirls are fit with the automated methods while 6$\%$ had no detectable radial expansion signatures (Swirls 16 and 17).
In terms of $m$, 24$\%$ of swirls had significant peaks at $m = 0$, 41$\%$ at $m = \pm 1$, and 9$\%$ at $m = \pm 2$ while 26$\%$ have no $m$ values.

\begin{figure*}
    \centering
\includegraphics[width=\textwidth]{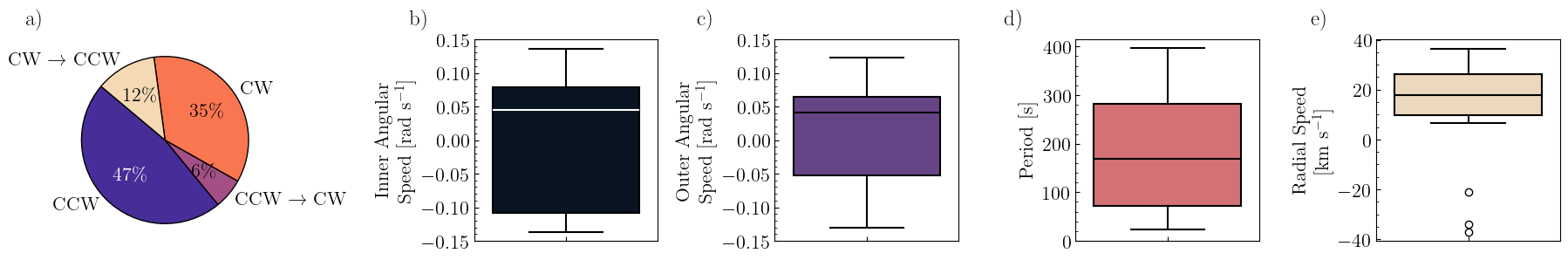}
    \caption{Statistics of the rotational dynamics of the observed swirls. a) The distribution of the overall rotation of the swirls. Clockwise motion is denoted as CW, and counterclockwise motion is denoted as CCW. b) The distribution of the angular speed computed near the swirl centre, within a radius of 0.24--0.32\,Mm. c) The distribution of the angular speed calculated away from the swirl centre, beyond a radius of 0.32\,Mm (dependent on swirl diameter). d) The distribution of the derived radial speeds. e) The distribution of the period associated with the derived radial speed.}
\label{fig:rotational_statistics}
\end{figure*}

The statistical results are presented in Fig.\,\ref{fig:rotational_statistics}.
The results show several notable features of the observed swirl behavior: (1) six swirls (18$\%$) show changes in rotational motion with temporal evolution; (2) on average, swirls show decreasing angular speed with increasing radial distance from the swirl centre; and (3) swirls show large radial expansion speeds reaching $v_{r} = \pm\,36.9$\,km\,s$^{-1}$.

We detect a modest difference between the swirls rotating entirely counterclockwise (47$\%$) versus clockwise (35$\%$).
Additionally, six swirls (18$\%$) display reversed rotational motion when measuring rotational dynamics in time.
Interestingly, the swirls in this group are mostly spiral-shaped or change morphologies from spiral-shaped to circular.
Two swirls showing changing morphologies that exhibit this rotational dynamic are part of a twin swirl system (Swirls 4 and 12).
As these swirls evolve and interact with their twin swirl, their rotation progressively aligns with their counterparts and the entire system.

We also note a moderate decrease in angular speed with increasing radial distance from the swirl centre, with a Pearson coefficient of 0.54.
We find a median angular speed of 0.045\,rad\,s$^{-1}$ with an interquartile range (IQR) of 0.034--0.153\,rad\,s$^{-1}$ near the swirl centre and a median angular speed of 0.04\,rad\,s$^{-1}$ (IQR: 0.024--0.092\,rad\,s$^{-1}$) near the swirl arms.
No significant correlation was detected between the angular speed and either swirl diameter or lifetime.
While the swirls show radial expansions speeds reaching $\pm$\,36.9\,km\,s$^{-1}$, the median radial speed is 17.7\,km\,s$^{-1}$ (IQR: 7.8--8.6\,km\,s$^{-1}$), where 75$\%$ of swirls have speeds less than or equal to 26.3\,km\,s$^{-1}$.
We also note that three swirls show negative radial speeds up to -36.9\,km\,s$^{-1}$ (Swirls 6, 21, and 22), which might indicate plasma traveling inwards or a contracting motion rather than expansion.
All three of these swirls have circular morphologies.
The associated average period of the swirls is 180\,$\pm$\,114\,s.

\subsection{Correlation with BPs}\label{subsec:photosphere_rotational_dynamics}

Both observations and numerical simulations suggest that BPs have matching rotary motions to their overlying chromospheric swirls \citep{2011_Shelyag, 2019_Shetye_Verwichte}, although this has only been investigated for a select few cases.
We explore the correlation between these two to assess whether the motions or properties of BPs can serve as a diagnostic signature for the formation of chromospheric swirls.
Summarized correlations can be found in Table\,\ref{tab:BP_stats_correlations}.

Out of the 41 tracked BPs, only 34$\%$ are present before the swirl's lifetime (ranging from 21\,s to 28\,min), with less than half of those still present following the end of the swirl's lifetime.
Most of the BPs (69$\%$) appear within 12\,s to 9\,min after the swirl becomes visible, and the majority (72$\%$) disappear during or before the swirl's lifetime ends.

We qualitatively find that the derived swirl rotation either matches the BP's dominant orientation (49$\%$) or the BP's overall turning direction along its trajectory (37$\%$), suggesting a connection between BP motions and swirl dynamics.
However, the derived swirl rotation matches neither variable in 15$\%$ of cases.
These dynamics can be seen when comparing the motions of BPs and swirls in Fig.\,\ref{fig:MBP_tracking_example}.
In particular, Swirl 30 exhibits a time-dependent rotation, transitioning from counterclockwise to clockwise, matching the BP's turning direction along its trajectory.

Moderate positive correlations were also observed between BP lifetimes and swirl diameters (Pearson coefficient of 0.55) and between BP angular speeds and swirl angular speeds with increasing radial distance from the swirl centre (Pearson coefficient of 0.46).
In contrast, no significant correlations were found between the number of associated BPs and either swirl diameter or swirl lifetime, nor between BP lifetimes and swirl lifetimes (Pearson coefficients < 0.25).

\section{Temporal Evolution and Formation of Swirl 18 by a Photospheric Vortex} \label{sec:swirl_evolution}

\begin{figure*}
\centering\includegraphics[width=\textwidth]{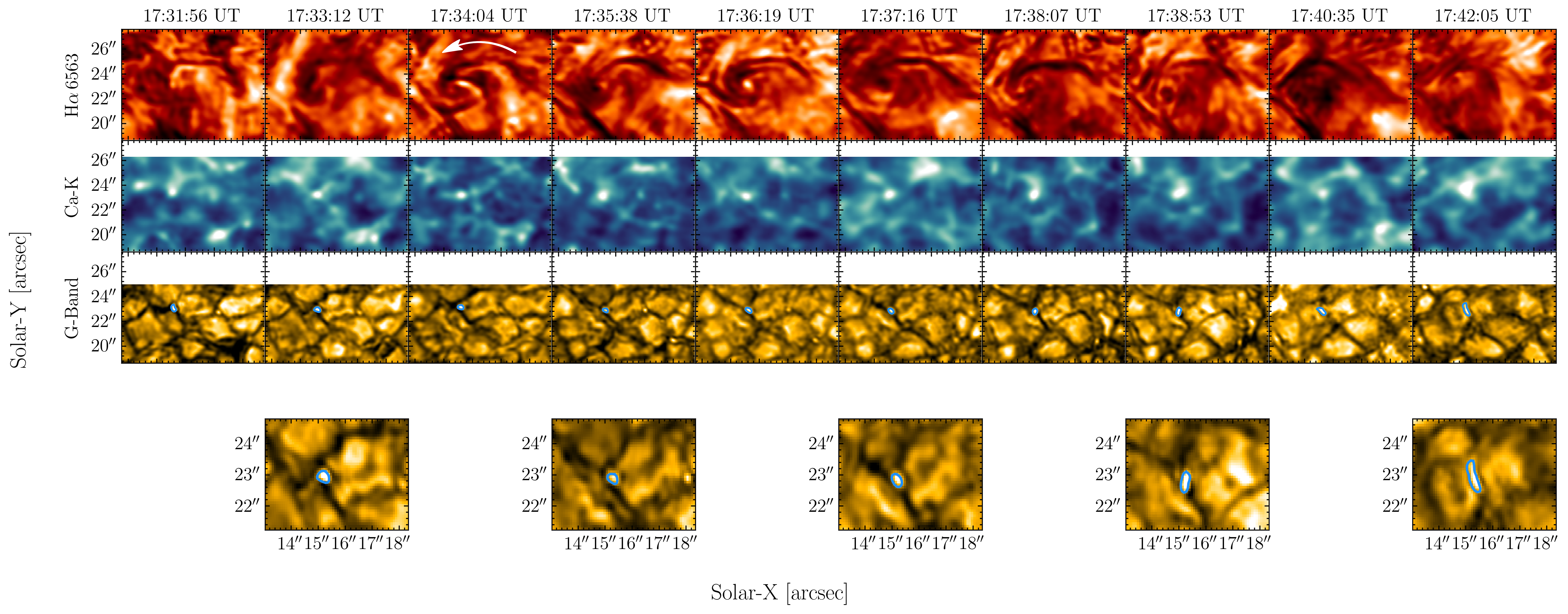}
    \caption{The temporal evolution of Swirl 18 mapped out from the low photosphere to the middle chromosphere from 17:31:56 to 17:42:05\,UT. White arrows denote the counterclockwise rotation of this large spiral-shaped swirl. The last row shows a zoomed-in G-Band sequence to highlight the BP's changes in orientation and morphology. An animation is available online.}
    \label{fig:swirl17_evolution}
\end{figure*}

We present observations of the interaction of a BP with a long-lasting photospheric vortex and the subsequent evolution and formation of a chromospheric swirl.
Swirl 18's well-defined signatures in both the photosphere and chromosphere make it a valuable case study for probing the underlying physical processes of chromospheric swirls despite its atypical nature.

Fig.\,\ref{fig:swirl17_evolution} shows the temporal sequence of Swirl 18 from 17:31:56 to 17:42:05\,UT, mapped from the low photosphere to mid-chromosphere. 
Swirl 18 exhibits well-defined spiral arms, a diameter of 7.4\,Mm, a lifetime of 10.5\,min, and a clear counterclockwise rotation ($\approx$\,0.1\,rad\,s$^{-1}$).
We derived a radial expansion speed of $\approx$\,33\,km\,s$^{-1}$ with a period of 311\,s.

The beginning of the H$\alpha$ evolutionary sequence is marked by the dark intensity imprint of the top spiral arm, which is stable and prominent throughout its lifetime.
As the swirl evolves, its central structure exhibits the most change.
The defined spiral structure that is noticeable at 17:34:04\,UT is only visible for less than a minute.
The centre also shows an alternating dark and bright spot from 17:34:04 to 17:37:08\,UT.
This central bright spot is co-spatial with the bright spot visible in Ca-K and the BP found in the lower photospheric wavelengths.
The co-spatial nature, along with the increasing size of the bright spot throughout the atmosphere, supports the idea that the chromospheric swirl is a coherent structure with a footpoint in the photosphere \citep{2014_Wedemeyer_Steiner}.
 
Swirl 18 is associated with a long-lasting BP, with an average diameter of 183\,km and a lifetime of 43\,min.
While BPs generally have short lifetimes, long-lived BPs have been detected previously, and these BPs are related to stable, long-lived magnetic field concentrations \citep{Nisenson_2003, 2014_Bodnarova_Utz_Rybak}.
The BP is visually resolved approximately 28\,min before the swirl's lifetime and disappears 3\,min after the swirl.
During its temporal evolution, the BP exhibits visual morphological changes as seen in Fig.\,\ref{fig:swirl17_evolution}, where it transitions from a compact, quasi-circular feature to an elongated feature.

When the swirl is at its most defined spiral shape, the BP is roughly compact in appearance with an eccentricity of 0.6 and a semi-major axis of 113\,km.
It is not until the end of the swirl's evolutionary sequence, where it loses its spiral definition, that the BP becomes noticeably elongated with an eccentricity of 0.92 and a semi-major axis of 206\,km.
The BP also exhibits a visible change in orientation, as seen in the zoomed-in G-Band sequences from 17:37:16 to 17:42:05\,UT.
The BP experiences orientation changes from $-75^{\circ}$ to $11^{\circ}$ to $-41^{\circ}$ from the vertical.
Analysis of its dominant orientation points to a predominantly counterclockwise motion.
Additionally, the BP's overall turning direction along its trajectory is predominantly counterclockwise (see Fig.\,\ref{fig:MBP_tracking_example}).
The change in orientation and turning direction matches the swirl's counterclockwise rotation.

\begin{figure*}
\centering
\includegraphics[width=\textwidth]{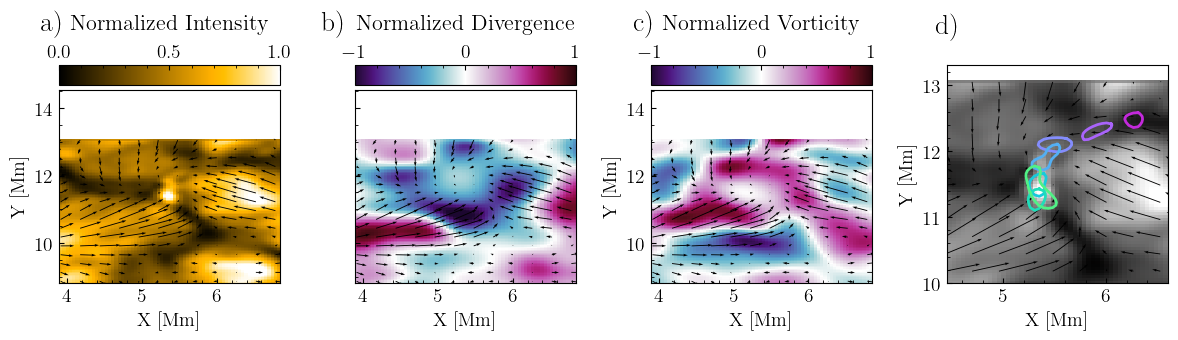}
    \caption{The inferred G-Band horizontal flow field derived from the LCT algorithm and BP evolution underneath Swirl 18. a) The horizontal flow field superimposed on the normalized G-Band intensity image averaged $\pm\,10$\,min around the swirl's lifetime. The BP is roughly in the centre of the photospheric vortex flow. b) The horizontal flow field superimposed on the normalized divergence field. c) The horizontal flow field superimposed on the normalized vorticity field. d) The normalized G-Band image from panel a) with the horizontal flow field and the BP's evolution and trajectory superimposed. The temporal evolution of the BP approximately every 7\,min is color-coded from purple to neon green from 17:03:25 to 17:42:00\,UT. Note that to capture the BP's final temporal changes, the neon green contour is only separated from the previous teal contour by 3.5\,min.}
\label{fig:swirl17_MBP_trajectory_flow}
\end{figure*}

To investigate the photospheric horizontal flow fields, we applied the local correlation tracking algorithm\,\citep[LCT;][]{1988_November_Simon} to the G-Band sequence after applying a subsonic filter to filter out $p$-modes.
The LCT algorithm uses a temporal window of 1\,min and an apodization window of 1\,arcsec.
The flow fields were then averaged over the entire duration of the G-Band time series (58.6\,min).
The inferred horizontal flow fields and the BP's trajectory are seen in Fig.\,\ref{fig:swirl17_MBP_trajectory_flow}.
While not shown, we qualitatively see the same results when applying the method to the 4170\,{\AA} sequence and simply averaging over the BP's lifetime.

\begin{figure*}
  \includegraphics[scale=0.43]{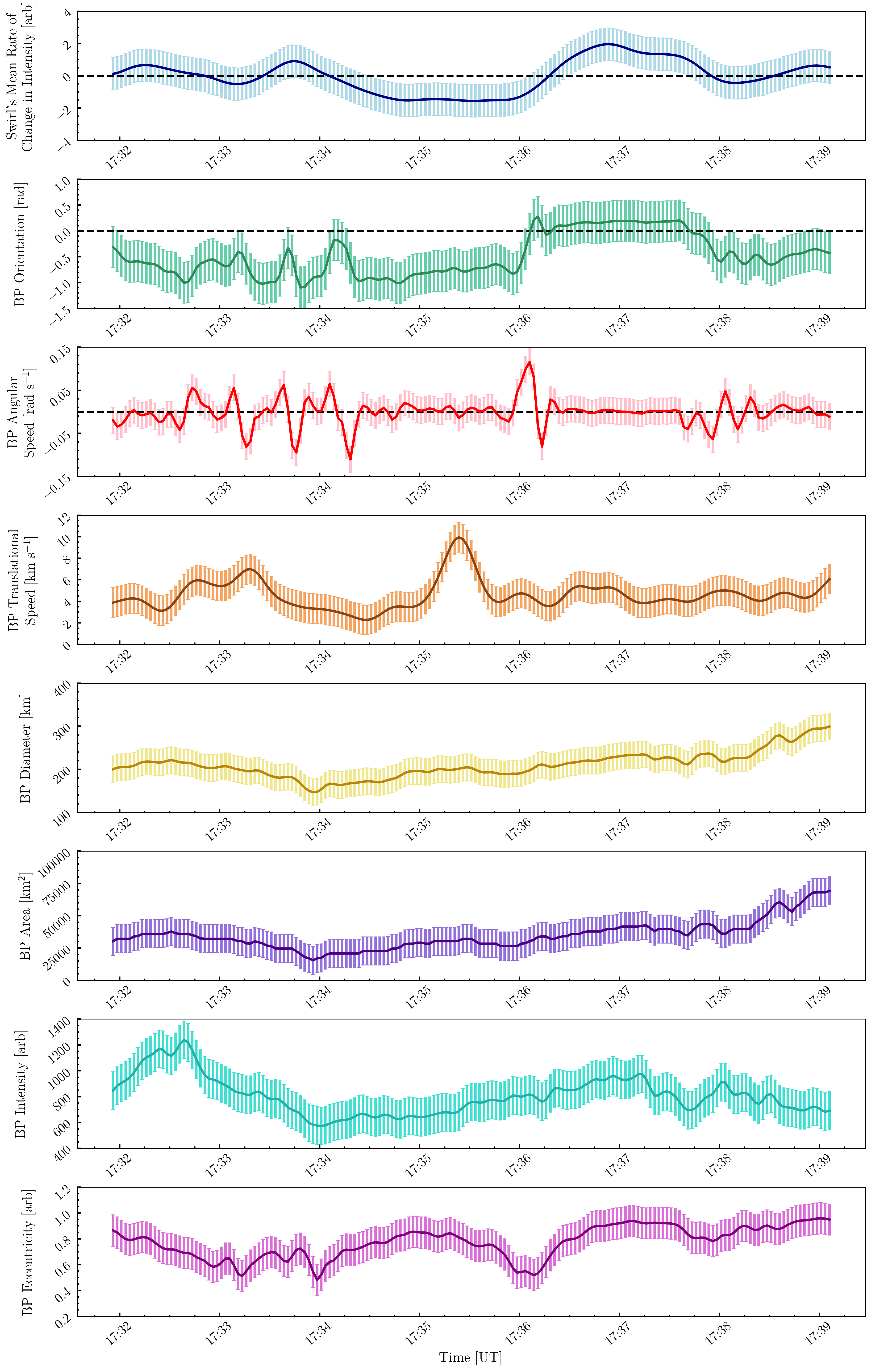}
    \caption{Temporal evolution of various BP properties and the mean rate of change in Swirl 18's intensity during the swirl's lifetime. The BP properties are derived from the tracking procedure in Section\,\ref{sec:photospheric_dynamics}. The swirl's mean rate of change in intensity is derived from averaging over all azimuthal angles and radii (Subsection\,\ref{subsec:chromosphere_rotational_dynamics}) and computing the accumulative integral of the resulting profile.}
    \label{fig:swirl17_BP_correlation}
\end{figure*}

We detect a photospheric vortex flow with a diameter of 2.4\,Mm situated over the intergranular lanes where the BP resides, which is co-spatial with the chromospheric swirl.
The photospheric vortex exhibits strong negative divergence and positive vorticity around BP, while the central portion shows near-zero divergence and vorticity (0.06\,s$^{-1}$ and 0.008\,s$^{-1}$, respectively).
For a rigid body rotation, this translates to an approximately 26\,min period.
The strong negative divergence indicates the location of a sink \citep{2010_Balmaceda} while the strong positive vorticity indicates a counterclockwise swirling motion \citep{1997_Simon_Weiss}.
This vortical flow pattern is characteristic of the ''bathtub effect`` \citep{1985_Nordlund}.
We find that the counterclockwise rotation of the vortical flow matches that of the chromospheric swirl and the BP.

As seen from Fig.\,\ref{fig:swirl17_MBP_trajectory_flow}d, the BP is displaced from its initial position and dragged toward the centre of the photospheric vortex flow.
This can be seen following its trajectory and visual inspection of the time series.
The BP is dragged towards the photospheric vortex after $\approx$\,21\,min, where, after an additional $\approx$\,7\,min, its location coincides with the centre of the photospheric vortex flow.
In 2D, this motion takes on the serpentine trajectory seen in Fig.\,\ref{fig:MBP_tracking_example}.
A brief time after the BP's position coincides with the centre of the vortex flow, the swirl's H$\alpha$ intensity imprint starts forming.
Within this vortex flow, the BP experiences a visible change in orientation and elongation, where it ultimately fragments. 
Analysis of the translational speed derived from the tracking procedure shows that the BP experiences an increase in speed from 4.6\,km\,s\,$^{-1}$ to 9.9\,km\,s\,$^{-1}$ as the swirl forms.

To investigate the coupling dynamics of this vortex system, we analyze the temporal relationship between various BP properties and the mean rate of change of intensity across the swirl by interpolating them onto the same time grid, as seen in Fig.\,\ref{fig:swirl17_BP_correlation}. 
The variation in swirl intensity mostly originates from the swirl's centre (see Fig.\,\ref{fig:swirl17_evolution}) and matches the temporal evolution of the BP's orientation.
We detect a time delay of $-42.5$\,s between the occurrence of the sudden change in the BP's orientation (17:36:11\,UT) and the peak of the swirl’s intensity variation (17:36:53\,UT).
We also note that the shape of the BP becomes more circular, i.e., smaller eccentricity, during times of significant rotation of the BP.
Finally, there is no temporal relation between the swirl intensity and the BP size or intensity.

\section{Discussion} \label{sec:discussion_and_conclusions}

\subsection{Swirl Properties}

While swirls have been primarily identified using H$\alpha$ narrowband spectroscopic imaging with a narrow bandpass filter \citep[e.g., CRISP H$\alpha$ observations employ a 0.061\,\AA{} bandpass filter;][]{2018_Tziotziou_Tsiropoula_etal,2019_Shetye_Verwichte}, HARDcam's H$\alpha$ line core has a broader bandpass filter of 0.25\,\AA{}.
Thus, the observed swirls are not velocity resolved, resulting in a different visual perspective because the intensity signatures contributing to their evolution in the wings of H$\alpha$, as observed in \citet{2019_Shetye_Verwichte}, are not separated in our observations.
Generally, most of our swirls display similar behavior to the \ion{Ca}{ii}\,8542\,{\AA} swirls described in \citet{2019_Shetye_Verwichte}, appearing as both dark and bright imprints.

We detect a moderate positive correlation between the swirl lifetimes and diameters, as seen in Fig.\,\ref{fig:diameter_lifetime_histogram_scatter}a.
The swirls have an average lifetime of 7.9\,$\pm$\,5\,min and an average diameter of 3.6\,$\pm$\,1\,Mm.
75$\%$ have lifetimes around 10.5\,min making them comparable to previous works \citep[e.g.,][]{2009_WedemeyerBohm_RouppevanderVoort, 2019_Shetye_Verwichte, 2022_Dakanalis}, and estimated diameters comparable to the H$\alpha$ vortex flow reported in \citet{2018_Tziotziou_Tsiropoula_etal} and \ion{Ca}{ii}\,8542\,{\AA} swirls reported in \citet{2012_Wedemeyer-Bohm_etal_Nature}.

We find that H$\alpha$ swirls tend to be small-scale, short-lived swirls.
This analysis agrees with the statistical study by \citet{2022_Dakanalis} using H$\alpha$-0.2\,{\AA} that high-resolution, high-cadence observations will result in a higher detection of small-scale swirls.
Our study supports that these phenomena are abundant in the quiet solar chromosphere, and investigations of swirls with instruments like the DST's high-cadence H$\alpha$ narrowband imager might shed more insight into their evolutionary nature.

The swirls show diverse appearances, although they can ultimately be grouped into morphological categories not too dissimilar from \citet{2013_Wedemeyer_etal}. 
We detect no preferential morphology associated with their H$\alpha$ imprints.
Although there are slightly more swirls classified under the complex category (38$\%$), this might be the result of visual bias or the effects of the broader H$\alpha$ bandpass filter (the intensity contributions from the wings are blended) given that \ion{Ca}{ii}\,8542\,{\AA} is nominally used to identify these structures.
These swirls exhibit well-defined structures in H$\alpha$, in contrast to those reported by \citet{2019_Shetye_Verwichte}, likely due to the factors above.
We cannot use Ca-K for a true lower chromospheric diagnostic to identify swirls, as its photospheric contribution drowns out the chromospheric signal \citep{2013_Keys_Mathioudakis}.

\subsection{Swirl-BP Dynamics}

While most swirls are associated with at least one co-spatial BP exhibiting a compact, quasi-circular morphology, we detect eight (23$\%$) swirls with no resolved BPs.
However, we cannot completely rule out that these swirls do not have BPs because the appearance of a BP as an intensity enhancement in G-Band depends on a variety of factors, such as the strength and inclination of the magnetic field \citep[see][]{Berger_2001, deWijn_2009}, and we cannot resolve features smaller than $\approx$100\,km.
Additionally, the lack of a resolved BP does not imply that these structures are not swirls nor connected to the magnetic field \citep[e.g.,][]{2007_Ishikawa, 2018_Tziotziou_Tsiropoula_etal}.
We also detect no noticeable characteristics between swirls with no resolved co-spatial BPs and those associated with at least one BP, confirming \citet{2019_Shetye_Verwichte}, suggesting that the BP's size is not an important factor for the appearance of a swirl.

Investigation of BP and swirl dynamics indicates that a dynamic coupling is evident between their motions, supporting both numerical simulations \citep{2014_Wedemeyer_Steiner} and observations \citep{2019_Shetye_Verwichte}.
We find three key results: (1) 69$\%$ of BPs appear following the appearance of a swirl; (2) BP global motions are significantly correlated with those of their overlying swirl; and (3) a moderate positive correlation exists between BP lifetimes and swirl diameters.

Most BPs appear as an intensity enhancement in G-Band following the visible appearance of a swirl (69$\%$), in the range of seconds to minutes.
This is in contrast to results by \citet{2019_Shetye_Verwichte} who found that BPs appear in \ion{Fe}{i}\,6302\,{\AA} Stokes-V channels at least 5\,min before the swirl.
The BP's delayed emergence as an intensity enhancement can likely be due to observational resolution limits, or the formation of the vortex systems, as described in \citet{2014_Wedemeyer_Steiner}, triggering the intensification of the magnetic field, which could result in their appearance \citep{2010_Danilovic_Schusslet_Solanki, 2018_Requerey}.

Additionally, while the shape of the BP's trajectory provides no direct connection to the appearance of a swirl, we find that in 85$\%$ of cases, the BP's global motions (overall turning trajectory and dominant orientation) match the rotation of the overlying swirl.
This supports the theoretical work of \citet{2011_Shelyag} and observational work of \citet{2019_Shetye_Verwichte}.
This can be understood as the BP's trajectory is affected by local horizontal motions present in the photosphere, driven by granulation and strong downflow events; however, the BP's global motions are likely driven by the same vortex flows powering the overlying swirl \citep{2014_Wedemeyer_Steiner}.

Our analysis also indicates that longer-lived BPs tend to be associated with larger swirls, and faster rotating BPs correspond to faster rotating swirl arms (Pearson coefficient of 0.55 and 0.46, respectively).
\citet{2014_Bodnarova_Utz_Rybak} find that BPs with longer lifetimes or those that travel farther are associated with more stable, long-living magnetic field concentrations.
It has also been observed that short-lived BPs are associated with weaker, more diffuse magnetic fields \citep{2015_Yang_Ji_Feng_Deng_Wang_Lin}.
This suggests that larger chromospheric swirls are associated with stronger, more coherent, long-living magnetic flux tubes rooted in the photosphere.

\subsection{Swirl Dynamics}

By examining the angularly integrated intensity as a function of azimuthal angle and radial distance versus time, we inferred the rotational and radial motions of the swirls.
The analysis indicates a modest difference with no statistical preferential rotation.
We find that 18$\%$ of swirls exhibit changing rotational motions with temporal evolution, with an equal amount classified as spiral and complex.
The swirls have angular speeds between $\pm$0.13\,rad\,s$^{-1}$ with median rotational periodicities of 140--157\,s ($\approx$ 0.04\,rad\,s$^{-1}$) and largely slow down with increasing radial distance from the swirl centre.
No significant correlation was detected between angular speed and either swirl diameter or lifetime, indicating that the rotational speed of swirls might be independent of their size or duration and suggesting that other factors govern their rotational dynamics.

These results are similar to those reported by \citet{2019_Shetye_Verwichte}, who analyzed swirls with angular speeds between 0.03--0.06\,rad\,s$^{-1}$ (105--209\,s).
Further work is needed to determine if the appearance of these H$\alpha$ swirls coincides with a reduction in the acoustic oscillations of the chromosphere, such as that reported by \citet{2019_Shetye_Verwichte}, and whether those acoustic oscillations or other wave modes, like those reported by \citet{2020_Murabito_Shetye_etal}, are influencing the apparent rotation of the swirls.

The swirls also show large apparent radial expansion speeds between $\pm$36.9\,km\,s$^{-1}$ with a median value of 17.2\,km\,s$^{-1}$, which is also not too dissimilar from previously reported results. 
This large median radial speed implies energetically significant dynamics occurring within the swirl. For a swirl with an average lifetime of 10\,min, such speeds would correspond to radial displacements exceeding tens of megameters, which is inconsistent with observations and numerical simulations of typical swirl diameters. This suggests that this measured radial expansion likely reflects the apparent motion of pattern features rather than continuous bulk plasma flow. The expansion may also be intermittent or localized, only contributing to short-lived, small-scale displacements.
Future work combining Doppler velocities from spectroscopic narrowband imagers with other diagnostics may clarify the relationship between apparent swirl motions and actual plasma flows.
From the automated methods, we note that the three swirls that exhibited negative radial expansion speeds were of a circular morphology.
Inwardly spiraling flows have been seen in small-scale vortices found in numerical simulations by \citet{2011_Moll}; however, more work is necessary to understand what this means for observations of chromospheric swirls.

\subsection{Swirl 18 and BP Temporal Dynamics}

Our analysis also investigated the interaction between a long-lived, large BP and a large, persistent photospheric vortex and a swirl's subsequent formation, amplification, and evolution in the middle chromosphere \citep{oana}.
While several of the swirls exhibited strong sinks or small photospheric vortices near their footpoints, we could not detect similar physical scenarios to that of Swirl 18.
Due to the choice of parameters and averaging over the entire time series, any short-lived photospheric vortices that might be associated with the shorter-lived swirls were effectively smoothed out.
Additional work will be done to explore the connection of swirls in the middle chromosphere to detectable vortical motions in the photosphere using open-source automated vortex identification tools such as the Automated Swirl Detection Algorithm \citep[ASDA;][]{2019_Liu_Nelson_Erdelyi, 2019_Liu} and the SWIRL code \citep{2022_Canivete_Cuissa_Steiner, 2024_Canivete_Cuissa_Steiner}.

We detected a 2.4\,Mm diameter photospheric vortex flow co-spatially underneath Swirl 18 with strong negative divergence and vorticity.
We observed the associated BP being dragged toward the centre of the photospheric vortex flow, changing appearance and orientation, where, after some time, an observable swirl imprint was identified in the middle chromosphere.
A similar result was shown by \citet{2024A&A...691A..37D} investigating the interaction of a BP with a photospheric intergranular flow and its relation to a chromospheric swirl, although with different dynamics.
Independent observations have also shown that the interactions between long-lived, large photospheric vortex flows and magnetic concentrations affect the dynamics and evolution of magnetic fields, producing strong downflows and amplification of the magnetic field \citep[see][]{2010_Balmaceda, 2010_Kitiashvili, 2017_Requerey,2018_Requerey,Keys_2020}.
As such, the interaction between the BP and the photospheric vortex flow probably played a significant role in the formation of Swirl 18, amplifying its evolution, as the visual intensity imprint of Swirl 18 is a direct result of its photospheric footpoint (the BP) coinciding with the centre of the photospheric vortex flow.
The visible effects on the morphology and orientation of the BP are consistent with numerical results by \citet{2012_Kitiashvili, Keys_2020} and observational results from \citet{2018_Requerey}, that when a BP is caught within a vortex flow, the associated flux tube becomes stable; however, once the vortical flow weakens, the BP eventually disperses, shows orientation changes, and fragments.
As we could only identify one case of such dynamics, the frequency at which this generation mechanism occurs is unknown.
However, the observational detection of this process underscores that chromospheric swirls are related to vortex motions generated in the photosphere on many scales, which have been discussed in numerous numerical simulations \citep[e.g.,][]{2010_Carlsson, 2012_Kitiashvili, 2020_Silva, 2021_Yadav}.

We detect a temporal correlation between the BP's angular speed, orientation, and shape, and the swirl's mean rate of change of intensity in time. 
This suggests that the BP's rotational motion drives dynamics in the overlying swirl.
The temporal correlation between the pulse in the BP's angular speed and the sudden change in orientation at 17:36:11\,UT, and the subsequent increase in the swirl's intensity at 17:36:53\,UT is consistent with an upward propagating torsional Alfvén wave.
We speculate that the rapid change in the BP's orientation acts as a driver, injecting torsional energy and exciting an Alfvén pulse that propagates along the magnetic field \citep[e.g.][]{Suzuki2005,Verth2010b,Fedun2011,Soler2017}.
This wave may be accompanied by compressions and rarefactions either due to its collective property due to the presence of magnetic twist and rotation \citep[e.g.][]{Vasheghani2017b}, or due to its nonlinear ponderomotive force \citep{Verwichte1999,Terradas2004,Vasheghani2012}. 
These density fluctuations can manifest as intensity fluctuations in the centre of the swirl.
The observed time lag of $-42.5$\,s between the occurrence of the sudden change in the BP's orientation and the maximum variation in swirl intensity implies a travel speed of 33\,km\,s$^{-1}$ over a height of 1.4\,Mm.
This value is not too dissimilar with the Alfvén speed of 22\,km\,s$^{-1}$ derived by \citet{2009_Jess_Mathioudakis_Erdelyi_Crockeet_etal}, who detected torsional Alfvén waves in a BP group, and comparable with an object with 1.5\,kG magnetic field strength assuming an electron density of $10^{16}$\,cm$^{-3}$.
This supports the idea that torsional Alfvén waves driven by BP rotation may play a central role in the formation and evolution of chromospheric swirls.
We do not find clear evidence for repetitive translational motions of the swirl centres in H$\alpha$, nor do we observe oscillatory patterns in radial velocity or intensity that would be indicative of kink or sausage modes, respectively. While kink and sausage modes have been reported in other chromospheric features \citep[e.g.,][]{2013_Morton, 2023_Jess_Jafarzadeh_keys_etal}, including swirls \citep[e.g.,][]{2019_Shetye_Verwichte, 2020_Murabito_Shetye_etal}, we do not detect their presence in our observations. This suggests that such modes are either absent or below the detection threshold imposed by the resolution, the limited lifetime of the observed swirls, or the diagnostics employed.

\section{Conclusions} \label{sec:conclusions}

This study presents the first detection of chromospheric swirls using high-resolution, multi-height imaging data from the DST's HARDcam and ROSA instruments.
We investigated the photospheric and chromospheric morphologies and dynamics of 34 swirls at quiet Sun disk centre, expanding the sample of swirl observations and investigations of swirl-BP dynamics.

Our analysis reveals a dynamic coupling between the motion of BPs and the overlying motions of chromospheric swirls, solidifying their connection, in agreement with \citet{2014_Wedemeyer_Steiner}.
A unique case study highlighted the formation and evolution of a large spiral-shaped chromospheric swirl above the interaction of its footpoint with a large photospheric vortex.
The observed time delay and temporal correlation between the sudden change in the BP's angular velocity and orientation and the swirl's maximum intensity variation highlight the BP's role in driving the motions of the overlying swirl.
This points to a signature of a torsional Alfvén wave being driven upwards, powering the swirl.

These findings reinforce that swirls are abundant, dynamic phenomena magnetically anchored in the photosphere to BPs.
DKIST’s unprecedented spatial resolution and multi-wavelength capabilities will enable the detection of finer, faster-evolving swirls and the resolution of substructures previously hidden at smaller scales, including that of BPs, which would shed more insight into their dynamics with swirls \citep{2024_Van_Kooten_Cranmer}.
Additionally, simulations by \citet{2023_Matsumoto_Kawabata_Katsukawa_etal, 2024_Kuniyoshi_Bose_Yokoyama} predict that highly twisted magnetic fields produce arc-like linear polarization signals that can be detected in strong chromospheric lines, providing a potential observational signature of the magnetic field and driving mechanism of these tornado-like structures. These structures could be observed with future DKIST polarimetric measurements.
These observations will be crucial for clarifying the magnetic and dynamic coupling of vortex flows across the lower solar atmosphere and quantifying their role in wave generation and energy transport. 
Future comparisons with numerical simulations \citep[e.g.,][]{2021_Battaglia_CaniveteCuissa_Calvo_etal, 2022_Finley_Brun_Carlsson_etal} will also be important for constraining physical interpretations.



\section*{Acknowledgements}

Data in this publication was obtained with the Dunn Solar Telescope facility, which is operated by New Mexico State University with funding support from the National Science Foundation (NSF) and the state of New Mexico.
O.V. acknowledges support by the COFFIES DSC Cooperative Agreement 80NSSC22M0162 and the New Mexico Space Grant Scholarship and Fellowship Program. 
The funding for the New Mexico Space Grant Scholarship and Fellowship Program is contained in the National Aeronautics and Space Administration (NASA) Cooperative Agreement 80NSSC20M0034. 
J.S. acknowledges funding from NSF grant \# 1936336.

\section*{Data Availability}
The ROSA and HARDcam datasets are publicly available on the Sunspot Solar Observatory Consortium (SSOC) Data Archive (\href{http://ssoc.nmsu.edu/}{http://ssoc.nmsu.edu/}).



\bibliographystyle{mnras}
\bibliography{refs} 




\appendix

\section{Observation Details}\label{sec:observation_table}
\begin{table*}
\caption{Overview of observations and spectral lines.\,\label{tab:dataset_overview}}  
\small
\centering          
\begin{tabular}{c l c c c c c c c }     
\hline\hline  
Dataset & Wavelength & Location & Mean & FOV & Start Time & Duration & Reconstructed & Exposure \\
{} & {} & {} & Scintillation & {} & [UT] & [min] & Cadence [s] & [ms] \\
\hline \\[-1.5ex]                   
   {} & ROSA G-Band & {} & {} & 53\arcsec\,$\times$\,52\arcsec  & 15:13:47 & 82.1 & 2.1 & 21\\
   10 November 2017   &   ROSA 4170 & (0.0\arcsec, 0.0\arcsec) & 0.79\arcsec & 53\arcsec\,$\times$\,52\arcsec &  15:13:47 & 82.1 & 2.1 & 21 \\
   {(DS1)} & ROSA Ca-K & {} & {} & 53\arcsec\,$\times$\,52\arcsec & 15:13:47 & 40.9 & 8.4 & 140\\
   {} & HARDcam H$\alpha$ & {} & {} & 166\arcsec\,$\times$\,165\arcsec & 15:13:47 & 102.6 & 3.2 & 50\\
\hline \\[-1.5ex]
   {} & ROSA G-Band & {} & {} & 54\arcsec\,$\times$\,53\arcsec & 16:05:30 & 68.7 & 4.2 &8\\
   11 April 2018 &  ROSA 4170 & (0.0\arcsec, 0.0\arcsec) & 1.06\arcsec & 53\arcsec\,$\times$\,54\arcsec & 16:05:30 & 68.7 & 4.2 & 10\\
   {(DS2)} & ROSA Ca-K & {} & {} &54\arcsec\,$\times$\,53\arcsec & 16:05:30 & 68.8 & 10.5 &120\\
   {} & HARDcam H$\alpha$ & {} & {} & 83\arcsec\,$\times$\,81\arcsec & 16:06:00 & 36.0 & 2.5 &40\\
   \hline \\[-1.5ex]
   {}  & ROSA G-Band & {} & {} &55\arcsec\,$\times$\,53\arcsec & 16:58:00 & 58.5 & 2.1 & 10 \\
   27 November 2018 & ROSA 4170 & (0.1\arcsec, 0.0\arcsec) & 0.77\arcsec & 55\arcsec\,$\times$\,54\arcsec & 16:57:54 & 58.6 & 2.1 &14\\
   {(DS3)} &  ROSA Ca-K & {} &  {} & 55\arcsec\,$\times$\,55\arcsec & 16:57:54  & 58.4 & 10.5 & 140\\
   {} &  HARDcam H$\alpha$ & {} & {} &81\arcsec\,$\times$\,81\arcsec  & 16:57:48 & 57.2 & 2.5 & 40\\
   \hline \\[-1.5ex]
    {} &  ROSA G-Band & {} &  {} & 53\arcsec\,$\times$\,54\arcsec & 16:30:57 & 64.6 & 4.2 & 14\\
    04 December 2018 & ROSA 4170 & (-24.0\arcsec, -5.4\arcsec) & 0.83\arcsec & 53\arcsec\,$\times$\,53\arcsec & 16:30:57 & 64.6 & 4.2 & 14\\
    {(DS4)} &  ROSA Ca-K & {} & {} & 56\arcsec\,$\times$\,56\arcsec & 16:30:57 & 7.6 & 12.6 & 140\\
    {} & HARDcam H$\alpha$ & {} & {} & 165\arcsec\,$\times$\,165\arcsec & 16:30:53 & 65.0 & 2.8 & 45\\
\hline                  
\end{tabular}
\end{table*}

Table\,\ref{tab:dataset_overview} summarizes the high-cadence, multi-wavelength observations obtained with the DST's ROSA and HARDcam imaging instruments. 
The datasets, referred to as DS1, DS2, DS3, and DS4, were recorded on 10 November 2017, 11 April 2018, 27 November 2018, and 04 December 2018, respectively. 
Each dataset targets a quiet-Sun region near disk centre, with slight variations in pointing coordinates.
These datasets were selected for their excellent seeing conditions and simultaneous multi-height coverage, enabling us to track chromospheric swirls and their underlying photospheric structures over extended time intervals. 

The observations include simultaneous image sequences from ROSA in three channels (G-Band, 4170\,\AA{}, and Ca-K) and narrowband imaging in the H$\alpha$ line core from HARDcam. 
All ROSA channels have a spatial sampling of 0{\farcs}06 per pixel, while all the HARDcam data have a spatial sampling of 0{\farcs}083 per pixel.
The reconstructed cadences vary between 2.1–12.6\,s across the different channels, depending on the exposure times and temporal averaging employed during post-processing.
Among the datasets, DS1 and DS4 provide the longest uninterrupted H$\alpha$ sequences, while DS2 has the shortest H$\alpha$ duration but includes full coverage in all ROSA channels. 
DS4 includes a shorter Ca-K sequence due to instrumental limitations but provides high-cadence coverage in H$\alpha$.
Alignment is shown in Fig.\,\ref{fig:alignment_check}.

\begin{figure*}
        \centering
\includegraphics[width=\textwidth]{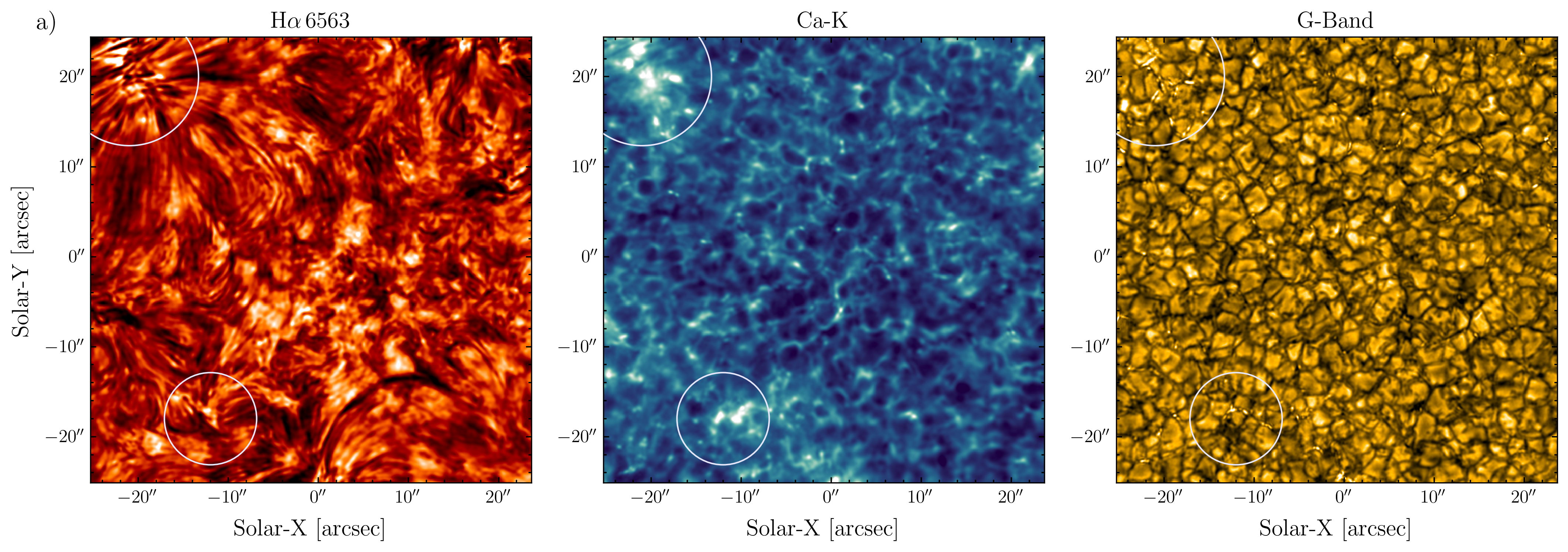}
\includegraphics[width=\textwidth]{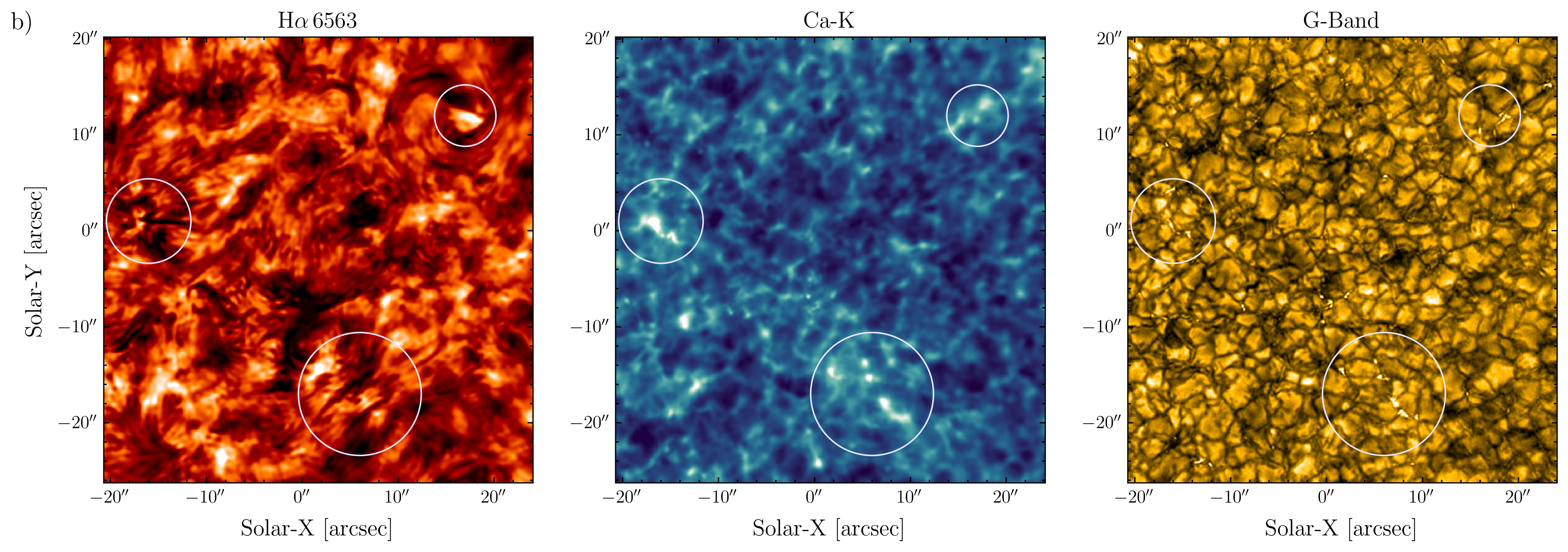}
                \caption{HARDcam (H$\alpha$) and ROSA (Ca-K and G-Band) co-temporal alignment across the four datasets: a) DS1, b) DS2, c) DS3, and d) DS4. Circles highlight bright, large-scale features used to ensure accurate co-alignment across the solar atmosphere.}\label{fig:alignment_check}
\end{figure*}

\begin{figure*}
        \centering
\includegraphics[width=\textwidth]{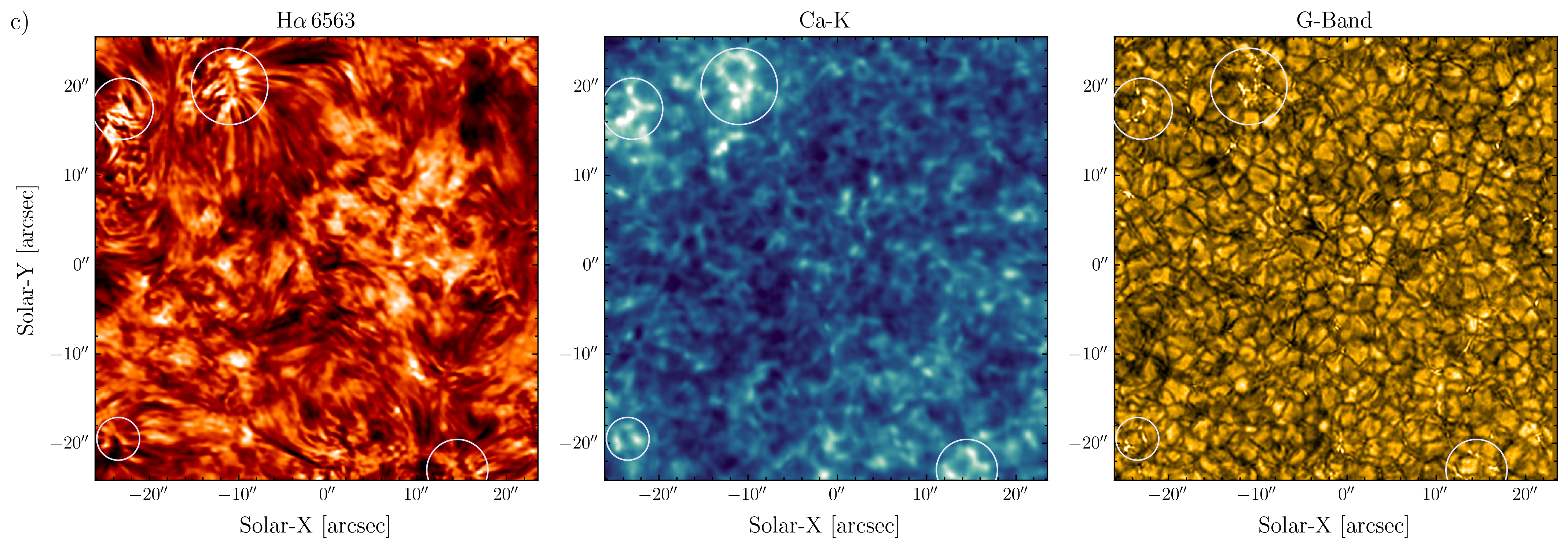}
\includegraphics[width=\textwidth]{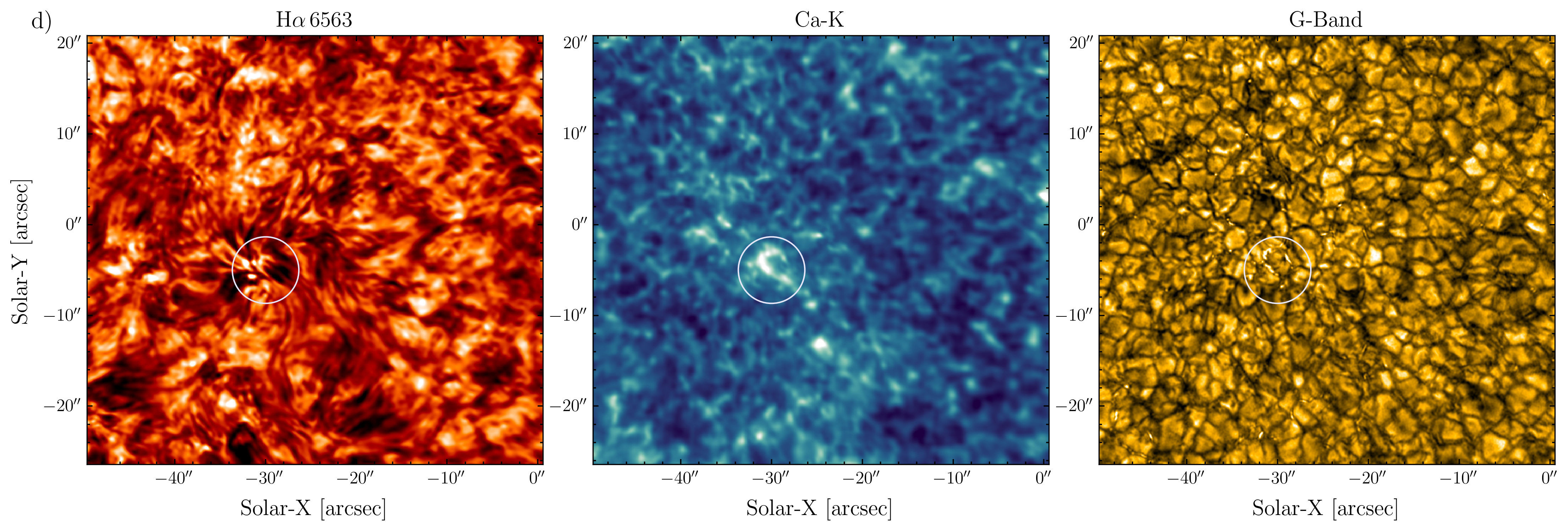}
                \contcaption{HARDcam (H$\alpha$) and ROSA (Ca-K and G-Band) co-temporal alignment across the four datasets: a) DS1, b) DS2, c) DS3, and d) DS4. Circles highlight bright, large-scale features used to ensure accurate co-alignment across the solar atmosphere.}
\end{figure*}

\section{Swirl Observations}\label{sec:swirl_properties}
The properties of the 34 chromospheric swirls observed with HARDcam (H$\alpha$\,6563).
We provide details regarding their approximate coordinates, lifetime in minutes based on their visible imprint, diameter in megameters, classified shape, rotation, angular speed, radial speed, and comments regarding their photospheric footpoints.
Clockwise rotation is denoted as CW, and counterclockwise rotation is denoted as CCW.

\begin{figure*}
    \centering
    \includegraphics[width=1.\textwidth]{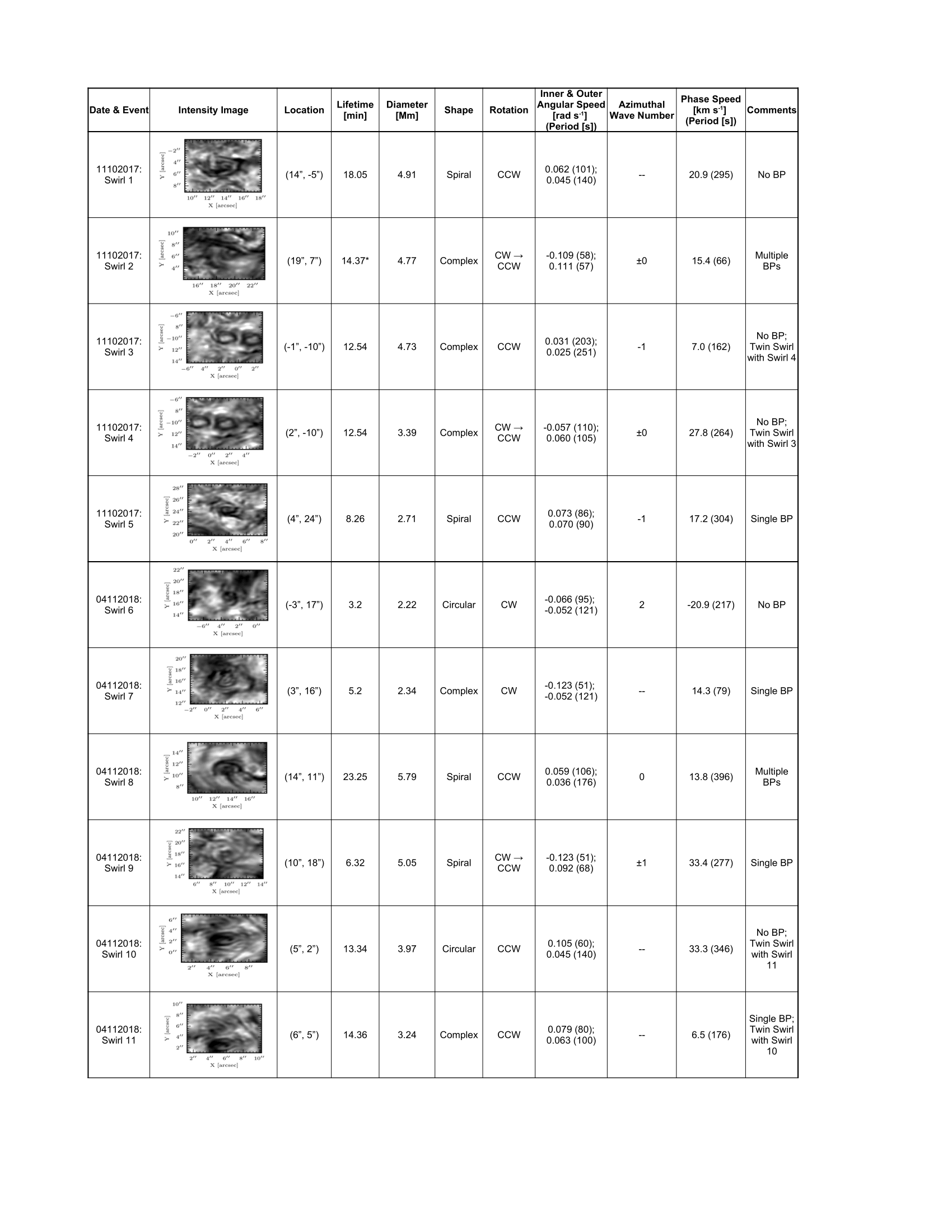}
        \caption{Chromospheric swirl observations with HARDcam's H$\alpha$. An asterisk next to the estimated lifetime indicates that these swirls likely started forming before the H$\alpha$ observations.}
\end{figure*}

\begin{figure*}
    \centering
        \includegraphics[width=1.\textwidth]{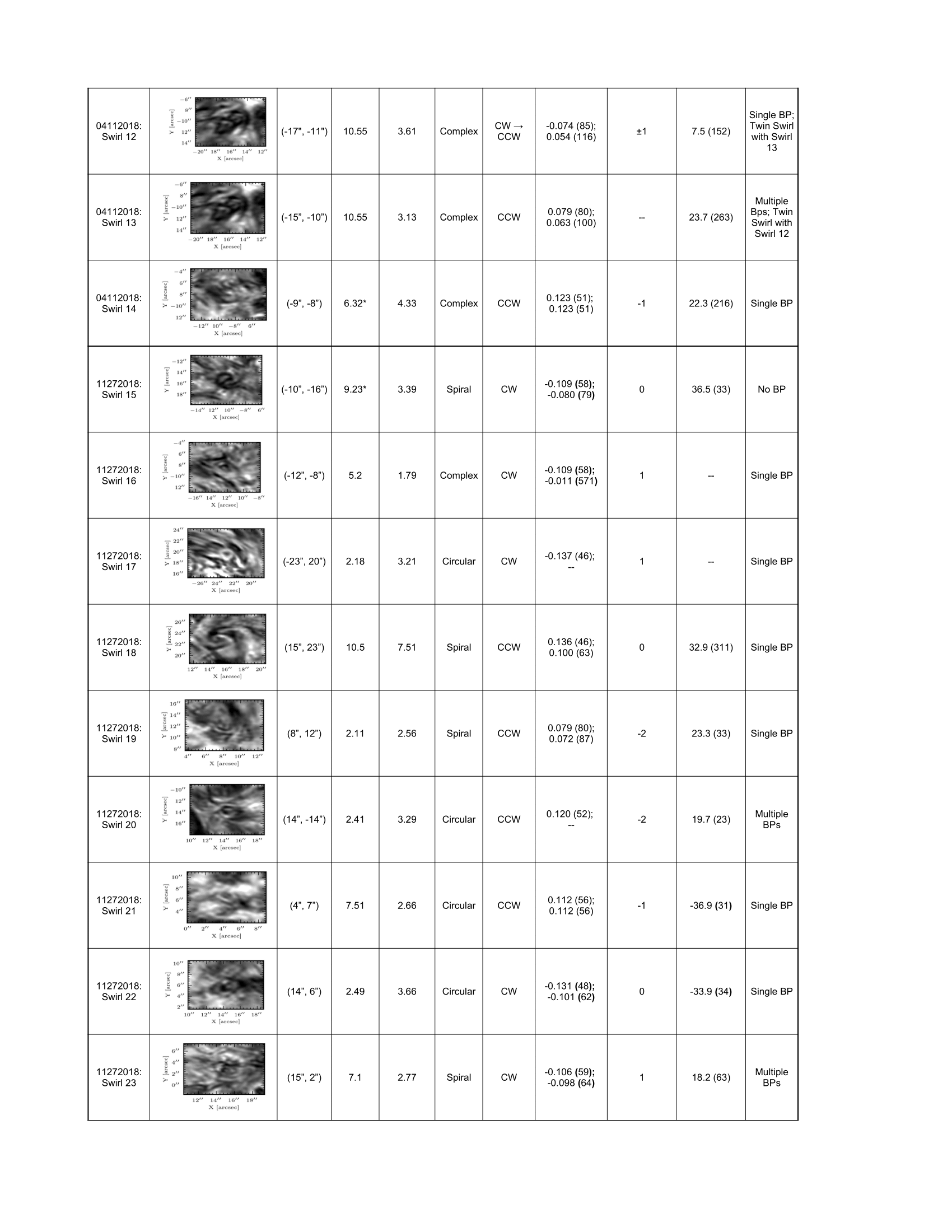}
                        \contcaption{Chromospheric swirl observations with HARDcam's H$\alpha$. An asterisk next to the estimated lifetime indicates that these swirls likely started forming before the H$\alpha$ observations.}

\end{figure*}

\begin{figure*}
    \centering
\includegraphics[width=1.0\textwidth]{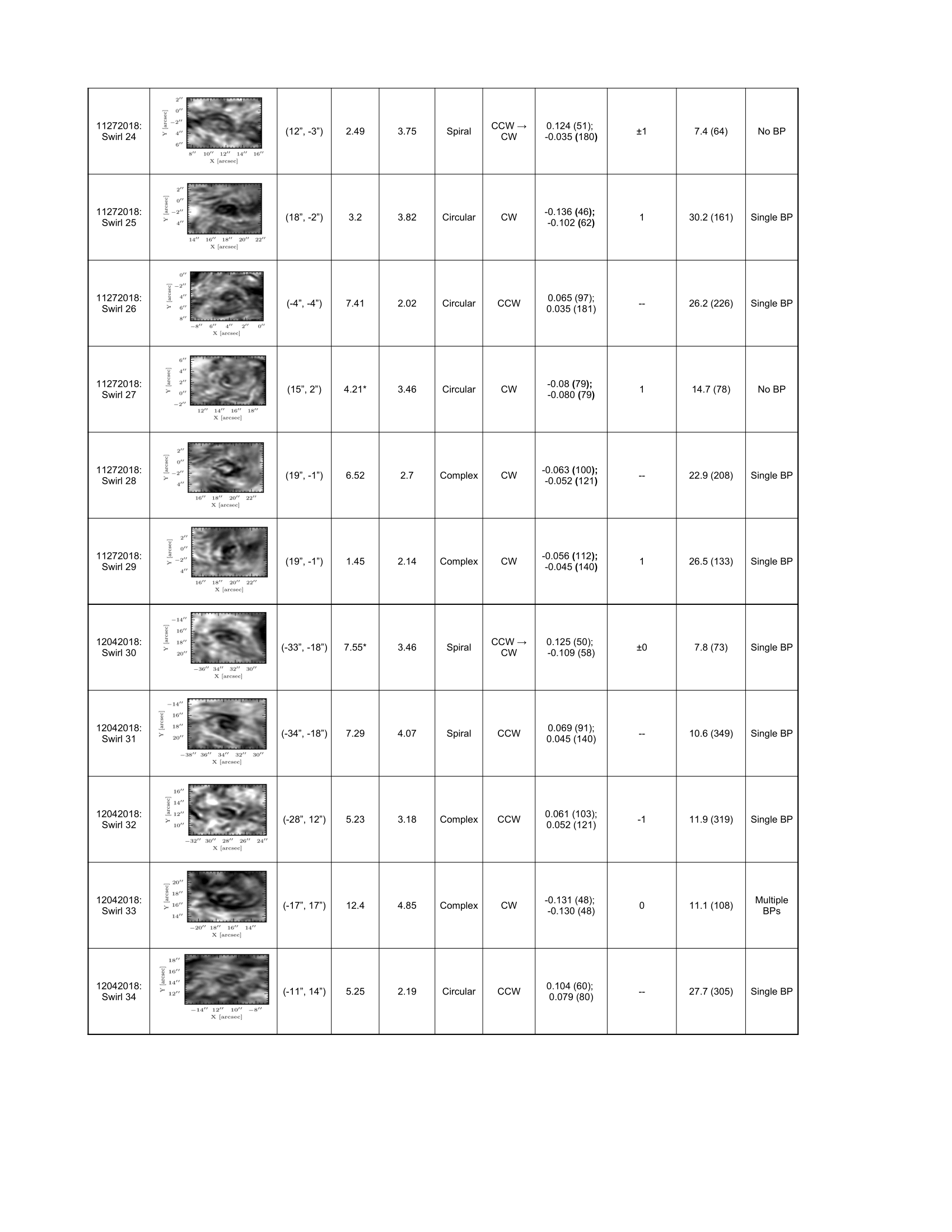}
                \contcaption{Chromospheric swirl observations with HARDcam's H$\alpha$. An asterisk next to the estimated lifetime indicates that these swirls likely started forming before the H$\alpha$ observations.}

\end{figure*}





\bsp	
\label{lastpage}
\end{document}